\documentstyle[aps,prbbib]{revtex}

\begin{document}

\title{
Langevin Simulation of Thermally Activated Magnetization
Reversal in Nanoscale Pillars
}

\author{
Gregory Brown$^{1}$ \and
M. A. Novotny$^{1}$ \and
Per Arne Rikvold$^{1,2}$
}

\address{
$^1$School of Computational Science and Information Technology,
Florida State University, Tallahassee, FL 32306-4120\\
$^2$Center for Materials Research and Technology and Department of Physics, 
Florida State University, Tallahassee, FL 32306-4350
}

\date{\today}
\maketitle

\begin{abstract}
Numerical solutions of the Landau-Lifshitz-Gilbert micromagnetic model
incorporating thermal fluctuations and dipole-dipole interactions
(calculated by the Fast Multipole Method) are presented for systems
composed of nanoscale iron pillars of dimension $9\,{\rm nm} \times
9\,{\rm nm} \times 150\,{\rm nm}$. Hysteresis loops generated under
sinusoidally varying fields are obtained, while the coercive field is
estimated to be $1979 \pm 14 \,{\rm Oe}$ using linear field sweeps at
$T=0\,{\rm K}$. Thermal effects are essential to the relaxation of
magnetization trapped in a metastable orientation, such as happens
after a rapid reversal of an external magnetic field less than the
coercive value. The distribution of switching times is compared to a
simple analytic theory that describes reversal with nucleation at the
ends of the nanomagnets. Results are also presented for arrays of
nanomagnets oriented perpendicular to a flat substrate. Even at a
separation of $300\,{\rm nm}$, where the field from neighboring
pillars is only $\sim 1\,{\rm Oe}$, the interactions have a
significant effect on the switching of the magnets.
\end{abstract}

\section{Introduction}
\label{sec:intro}

Several emerging technologies will incorporate fabricated magnets that
are small enough to contain only a single magnetic domain. These
nanoscale magnets will be essential for smaller components, lower
power consumption, and completely new applications in fields such as
information storage, integrated circuits, sensor technology, and
microelectromechanical systems. Successful implementation of these
technologies requires a fundamental understanding of the dynamics of
the internal magnetic structure of the nanoscale magnets on time
scales ranging from the nanoseconds associated with gigahertz
applications to the years over which magnetic information storage must
be stable.

An essential factor in many of these applications, especially in
information storage, is the free-energy barrier that separates two
antiparallel orientations of the magnetization. This free-energy
barrier can be surmounted using thermal energy momentarily
``borrowed'' from the surroundings, and often device engineers strive
for barriers of at least $40 k_{\rm B}T$ to make this thermal
bit-flipping a rare event. Here $T$ is the absolute temperature and
$k_{\rm B}$ is Boltzman's constant. The particle can be magnetizes in
a specified direction by applying a field strong enough to remove the
free-energy barrier.  The smallest field sufficient to do this at zero
temperature is called the coercive field $H_{\rm c}.$ Fields smaller
than the coercive field cannot change the magnetization orientation
deterministically, but they do cause lower free-energy barriers and
make thermal crossing more probable. In fact, hybrid recording
\cite{RUIGROK} uses lower-than-coercive fields and thermal barrier
crossing to write data in high-coercivity magnetic media.

Two examples of particular thermally activated barrier crossings are
shown in Fig.~\ref{fig:switchpicture}. Red is associated with the
magnetization orientation antiparallel to the applied field, which is
metastable. Blue is associated with the parallel magnetization, which
is thermodynamically stable. Green and yellow are intermediate between
the two orientations. For times before those shown in the figure, the
magnet stays in the metastable state with end caps of non-uniform
magnetization induced by pole avoidance. Eventually thermal
fluctuations in the end caps carry the system over the free-energy
barrier. After the barrier crossing, the entire magnet changes quickly
to the stable equilibrium orientation.

Often, nanoscale magnets are assumed to be uniformly magnetized
bodies. For instance, Garc{\`i}a-Palacios and L{\'a}zaro \cite{GARCIA}
considered the stochastic trajectories of isolated magnetic moments
using the Landau-Lifshitz-Gilbert equation subject only to applied
fields and a uniaxial anisotropy, which supplied the free-energy
barrier.  They measured the susceptibility of the magnetization in
small sinusoidal probe fields. Multiple crossing of the free-energy
barrier, a frequently noted consequence of the gyromagnetic motion in
single-spin models, was also observed for their uniform-magnetization
model.

When at least one dimension of the nanomagnet is greater than the
exchange length of the material, nonuniform reversal modes become
energetically possible. To consider such nonuniform magnetization
reversal we use micromagnetic simulation with a large number of points
inside the magnet. Specifically, nanomagnets with an aspect ratio of
approximately $17$ are considered throughout this work, and are 
referred to as pillars.  The numerical approach is discussed in
Sec.~\ref{sec:model}, with further details found in the
appendices. Results for the estimation of the coercive field and the
statistics of magnetization switching are discussed in
Sec.~\ref{sec:pillar}. A simple model for nonuniform switching is
discussed in Sec.~\ref{sec:1dpillar}. Section~\ref{sec:array}
considers the effect of interactions between pillars in a
one-dimensional array. A brief summary of the results is given in
Sec.~\ref{sec:summary}. Some preliminary results of this study were
presented in Refs.~\onlinecite{JAP2000,ASI,MMM01}.

\section{Numerical Details}
\label{sec:model}

The basic approach of micromagnetic modeling is to consider a system of
coarse-grained magnetization vectors
${\bf\tilde{M}}({\bf\tilde{r}}_i)$, with ${\bf\tilde{r}_i}$ indicating
the location in space of the $i$-th spin. The vectors are assumed to
have a fixed magnitude $M_s$, corresponding to the bulk saturation
magnetization density of the material. This is a valid approximation
for temperatures well below the Curie temperature. \cite{GARANIN}  The
evolution of each magnetization vector is governed by the damped
precessional motion given by the Landau-Lifshitz-Gilbert (LLG) equation
\cite{BROWN,AHARONI}
\begin{equation}
\label{eq:llg}
\frac{ {d} {\bf\tilde{M}}({\bf\tilde{r}}_i) }
     { {d} \tilde{t} }
 =
   \frac{ \gamma_0 }
        { 1+\alpha^2 }
   {\bf\tilde{M}}({\bf\tilde{r}}_i)
 \times
 \left (
   {\bf\tilde{H}}({\bf\tilde{r}}_i)
  -\frac{\alpha}{M_s} {\bf\tilde{M}}({\bf\tilde{r}}_i) \times 
                      {\bf\tilde{H}}({\bf\tilde{r}}_i)
 \right )
\;
\end{equation}
where the electron gyromagnetic ratio is $\gamma_0 = 1.76 \times 10^7
{\rm Hz/Oe}$, \cite{AHARONI} and $\alpha$ is a phenomenological
damping parameter. For the sign of the undamped-precession term we
follow the convention of Brown. \cite{BROWN} Here a tilde is used to
distinguish dimensional quantities from their dimensionless
counterparts, which will be introduced later.

The dynamics are controlled by ${\bf\tilde{H}}({\bf\tilde{r}}_i)$, the
local field at the $i$-th position, which, in general, is
different at each lattice site.  The local field mediates all of the
interactions in the system with the contributions combined via linear
superposition,
\begin{equation}
\label{eq:hterms}
{\bf\tilde{H}}({\bf\tilde{r}}_i) = 
  {\bf\tilde{H}}_{\rm z}({\bf\tilde{r}}_i) +
  {\bf\tilde{H}}_{\rm e}({\bf\tilde{r}}_i) +
  {\bf\tilde{H}}_{\rm d}({\bf\tilde{r}}_i) +
  {\bf\tilde{H}}_{\rm a}({\bf\tilde{r}}_i) +
  {\bf\tilde{H}}_{\rm n}({\bf\tilde{r}}_i)
\;,
\end{equation}
where 
${\bf\tilde{H}}_{\rm z}({\bf\tilde{r}}_i)$ represents the
externally applied field (Zeeman term), 
${\bf\tilde{H}}_{\rm e}({\bf\tilde{r}}_i)$ is due to exchange effects,
${\bf\tilde{H}}_{\rm d}({\bf\tilde{r}}_i)$ is the dipole field,
${\bf\tilde{H}}_{\rm a}({\bf\tilde{r}}_i)$ is due to crystalline
anisotropy, and
${\bf\tilde{H}}_{\rm n}({\bf\tilde{r}}_i)$ is the random field induced
by thermal noise. \cite{BROW63} The estimates used for each of
these fields are given below, except for 
${\bf\tilde{H}}_{\rm a}({\bf\tilde{r}}_i)$, which we have taken to
be zero. 
In this article, a $\sim$ emphasizes that the symbol represents a
quantity with units, the corresponding dimensionless quantity is
written without the tilde. Material parameters, such as $M_s,$ always
have units when appropriate.

We have chosen material parameters to match those of bulk iron; a
summary of all parameters used here is given in Table~1. The
saturation magnetization density is $M_s = 1700 \, {\rm emu/cm}^3,$
\cite{CULLITY} while the exchange length, the length over which
${\bf\tilde{M}}$ can change appreciably, is $l_{\rm e} = 3.6 \, {\rm
nm}.$ \cite{WIRTH98,EXCHANGE} The damping parameter $\alpha$ has proven
difficult to measure or estimate from {\em ab initio} considerations,
and may even depend on numeric details. \cite{FENG}  Here we have
chosen $\alpha=0.1$ to represent the under-damped behavior usually
assumed to exist in nanoscale magnets. In the other extreme, the
over-damped limit, the gyromagnetic motion intrinsic to the LLG
equation is suppressed, and the system can be more efficiently
simulated using Monte Carlo methods. \cite{NOWAK,NOWAK-BOOK}

The magnetization was discretized on a cubic lattice with
discretization length $\Delta \tilde{r} = 1.5 \, {\rm nm}$ chosen to
give sufficient resolution across the cross section of the pillar. The
time discretization of $\Delta \tilde{t} = 5 \times 10^{-14} \, {\rm
s}$ was chosen to avoid numerical instabilities in the
integration. The details of the integration scheme are given
below. The equation of motion was cast into dimensionless quantities
by considering the rescaled length ${\bf r}={\bf\tilde{r}}/l_{\rm e},$
time $t=\gamma_0 M_s \tilde{t},$ field ${\bf H}={\bf\tilde{H}}/M_s,$
and magnetization density ${\bf M}={\bf\tilde{M}}/M_s.$ This rescaling
is facilitated in cgs units, \cite{CULLITY} where magnetic fields can
be normalized by the magnetization density without the inclusion of
any constants.

In a continuum model, exchange represents local differences in the
alignment of the magnetization. The contribution to the local field
from exchange interactions can be approximated by $l_{\rm e}^2 \nabla^2
{\bf\tilde{M}}({\bf\tilde{r}})$\cite{AHARONI} which has been
implemented in the simulations by
\begin{equation}
\label{eq:exchange}
{\bf\tilde{H}}_{\rm e}({\bf\tilde{r}}_i) = 
\left( \frac{ l_{\rm e} }{\Delta \tilde{r}} \right)^2
\left(
-6 {\bf\tilde{M}}({\bf\tilde{r}}_i)
+\sum_{|{\bf\tilde{d}}|=\Delta\tilde{r}}
      {\bf\tilde{M}}({\bf\tilde{r}}_i+{\bf\tilde{d}})
\right)
\;,
\end{equation}
where the summation is over the six nearest neighbors of
${\bf\tilde{r}}_i.$
Here the exchange length is defined in terms of the exchange energy \cite{ARROTT}
$E_{\rm e}=-(\l_e^2/2)\int d {\bf r}\, {\bf \tilde{M}} 
\cdot {\bf \nabla}^2 {\bf \tilde{M}}.$
The dimensionless form of the exchange contribution to the local field
is
\begin{equation}
\label{eq:rexchange}
{\bf H}_{\rm e}({\bf r}_i) = 
\left(\frac{1}{\Delta r}\right)^2
  \left( 
     -6 {\bf M}({\bf r}_i)  + 
     \sum_{|{\bf d}|=\Delta r} {\bf M}({\bf r}_i+{\bf d}) 
   \right)
\;.
\end{equation}

The computationally most intensive part of determining the local field
is finding the contribution from dipole-dipole interactions,
${\bf{H}}_{\rm d}({\bf{r}})$.  Efficient calculation, which uses the
Fast Multipole Method (FMM), \cite{GREENGARD} is quite involved and is
discussed in Appendix 1.

Thermal fluctuations are included in the LLG equation of motion by
inclusion in the superposition of a random field ${\bf\tilde{H}}_{\rm
n}({\bf\tilde{r}})$ with Gaussian fluctuations whose first moments are
zero and whose second moments obey the fluctuation-dissipation
relation \cite{BROW63}
\begin{equation}
\label{eq:fluctuation}
\langle
 {\tilde{H}}_{{\rm n} \mu}({\bf\tilde{r}}_i,\tilde{t}) 
 {\tilde{H}}_{{\rm n} \mu'}({\bf\tilde{r}}_i',\tilde{t}')
\rangle
=   \frac{2 \alpha k_{\rm B}T}{\gamma_0 M_s V}
    \delta\left(\tilde{t}-\tilde{t}'\right)
    \delta_{\mu,\mu'}
    \delta_{i,i'}
\;,
\end{equation}
where ${\tilde{H}}_{{\rm n} \mu}$ indicates one of the Cartesian
components of ${\tilde{\bf H}}_{\rm n}.$ Here $V=(\Delta\tilde{r})^3$
is the discretization volume of the numerical implementation and
$\delta_{\mu,\mu'}$ is the Kronecker delta representing the
orthogonality of Cartesian coordinates. This result was derived for
isolated particles \cite{BROW63}, and interactions between the
discretization volumes can be important. \cite{SAFANOV} To make the
simulations tractable, the effect of interactions on the thermal noise
has been neglected. Here $\delta(\tilde{t}-\tilde{t}')$ is the Dirac
delta function, and its dimensions $[\delta(t-t')]=1/{\rm s}$ are
important. After the transformation to dimensionless quantities, the
fluctuation-dissipation result is 
\begin{equation}
\label{eq:rfluctuation} 
\langle {{H}}_{{\rm n} \mu}({\bf{r}}_i,{t})
{{H}}_{{\rm n} \mu'}({\bf{r}}_i',{t}') \rangle = \epsilon
\delta\left({t}-{t}'\right) \delta_{\mu,\mu'} \delta_{i,i'}
\;,
\end{equation} 
with the dimensionless strength of the stochastic field
given by 
\begin{equation} 
\label{eq:epsilon} 
\epsilon = \frac{2 \alpha
k_{\rm B}T }{M_{\rm S}^2 V} 
\;, 
\end{equation} 
where the untransformed
discretization volume and saturation magnetization are both used in cgs
units to yield the dimensionless result.  The dimensionless form of the
LLG equation is 
\begin{equation} 
\label{eq:rllg} 
\frac{{d}{\bf M}({\bf
r}_i)}{{d}t} = \frac{1}{1+\alpha^2}{\bf M}({\bf r}_i) \times \left[
  {\bf H}({\bf r}_i) - {\alpha} {\bf M}({\bf r}_i) \times {\bf H}({\bf
  r}_i) \right] 
\;.  
\end{equation} 
This dimensionless stochastic
differential equation is used for the numerical integration.

Since the LLG equation conserves the magnitude of the magnetization density,
each integration step amounts to a rotation. The size of the
tangential displacement at each integration step is given by
\begin{equation}
\label{eq:define:R}
{\bf R}({\bf r}_i,t) = 
\frac{1}{1+\alpha^2} {\bf M}({\bf r}_i,t) \times
\left[
({\bf I}({\bf r}_i,t) 
- \alpha {\bf M}({\bf r}_i,t) \times {\bf I}({\bf r}_i,t)
\right]
\;,
\end{equation}
where ${\bf I}$ is the ``impulse'' over the integration step defined below.
The magnetization after the integration step is given by
\begin{equation}
\label{eq:mnew}
{\bf M}({\bf r}_i,t+\Delta t) =
\frac{{\bf M}({\bf r}_i,t) + {\bf R}({\bf r}_i,t)}
     {\sqrt{1+{\bf R}^2({\bf r}_i,t)}}
\;,
\end{equation}
which ensures the conservation of unit magnitude for ${\bf M}$.
If only the deterministic dynamics were important a high-order
integration method could be used, but to correctly include the thermal
noise a lower-order method must be used.  Using first-order Euler
integration, the impulse including the fluctuating field is
\begin{equation} \label{eq:impulse}
{\bf I}({\bf r}_i,t) =
\left[
  {\bf H}_{\rm a}({\bf r}_i,t) + {\bf H}_{\rm d}({\bf r}_i,t)
 +{\bf H}_{\rm e}({\bf r}_i,t)
\right] \, \Delta t \,
+ \sqrt{\epsilon \, \Delta t} \, {\bf g}({\bf r}_i,t)
\;,
\end{equation}
where ${\bf g}({\bf r},t)$ is a random vector with each component chosen
independently from a Gaussian distribution of zero mean and unit
variance. This result, including the $\sqrt{\Delta t}$ contribution
characteristic of integration of stochastic processes, is explained in
Appendix~2.

\section{Isolated Nanomagnets}

\label{sec:pillar}

The numerical micromagnetic simulation methods described in the
previous section have been applied to model nanoscale magnets inspired
by those recently fabricated using a combination of chemical vapor
deposition and scanning tunneling microscopy techniques by Wirth, {\em
et al.\/} \cite{WIRTH98,WIRTH99} The technique has been used to
produce arrays of nanoscale magnetic particles with diameters down to
about $10$~{\rm nm} and lengths from $50$ to $250$~{\rm nm}. The
reversal of pillars with diameters more than twice the exchange
length, such as these, have been found numerically to proceed by a
mode where the magnetization is not constant across the diameter of
the pillar. \cite{HINZKE,NOWAK2,BERTRAM} The model particles considered here
are rectangular prisms $9$~{\rm nm}~$\times$~$9$~{\rm
nm}~$\times$~$150$~{\rm nm}, which require $N=4949$ sites on the
computational lattice. The long axis is taken as the $z$-axis, and the
cross-sectional area is chosen comparable to that for a $10$~{\rm nm}
diameter pillar. The external field is always applied along the
$z$-axis -- the easy axis induced by shape anisotropy. As mentioned in
Sec.~\ref{sec:model}, the material parameters were chosen to match
those of bulk iron. These are consistent with measurements on the
experimentally produced iron nanoparticles. \cite{WIRTH98,EXCHANGE}

\subsection{Coercive Field}

\label{sec:pillarcoerc}

\begin{figure}
\vskip 2.5in
\includegraphics{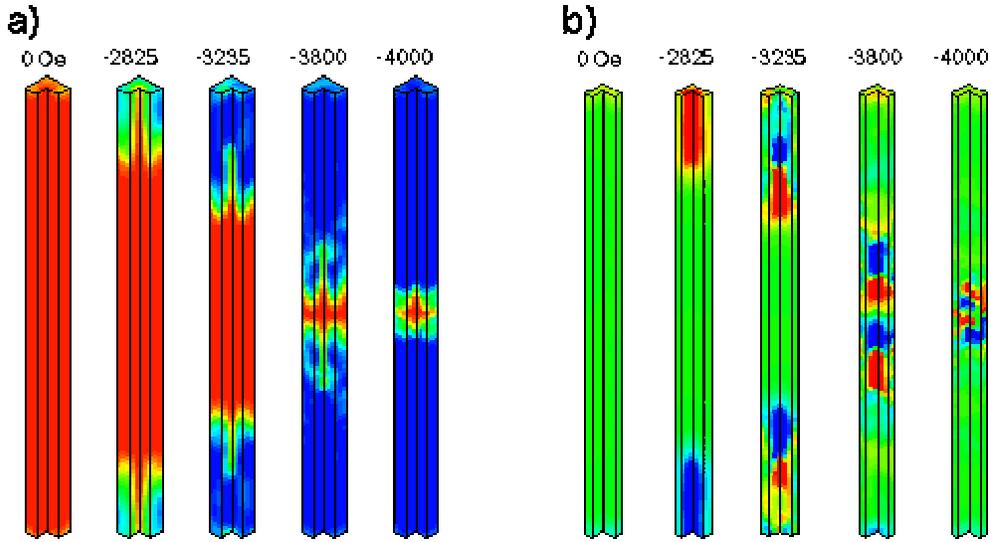}
\caption[] {Magnetization reversal during a hysteresis loop of
$1\,{\rm ns}$ at $T=0\,{\rm K}$. (a) The $z$-component of the
magnetization with the same color scale as in Fig.~1 (b) The magnitude
of the curl, $|{\bf C}|$, with the sign taken from the value of
$C_z$. Red represents positive curl, blue negative curl, and green
zero curl. The fields shown correspond to times $0.25$, $0.375$,
$0.405$, $0.45$, and $0.5$~ns, respectively.}
\label{fig:hystpicture} \end{figure}

Images of the $z$-component of the magnetization, $M_z$, for such a
particle as described above are shown in Fig.~\ref{fig:hystpicture}(a)
for several fields during a hysteresis-loop simulation (in a
one-quarter cut-away view). The color is a linear scale of $M_z,$
shown in the legend of Fig.~\ref{fig:switchpicture} with the most
negative values at the bottom and the most positive values at the
top. For hysteresis-loop simulations, the field is always oriented
along the $z$ axis. Starting from its most positive extreme, the value
is varied sinusoidally. For the simulation shown in
Fig.~\ref{fig:hystpicture} a period of one nanosecond, a field
amplitude of $4000\,{\rm Oe}$, and a temperature of $0\,{\rm K}$ were
selected. The response of the initially upward-directed magnetization
after the applied field is oriented downward can be clearly
seen. First, end caps associated with pole avoidance form at both
ends. These end caps are regions of high curl in the magnetization
density, ${\bf C}({\bf r})={\bf\nabla} \times {\bf M}({\bf r})$. The
end caps are obvious in Fig.~\ref{fig:hystpicture}(b), where ${\rm
sign} \left( C_z \left( {\bf r}, t \right) \right) | {\bf C} \left( r,
t \right) |$ is shown on a linear color scale, so that right
(left)-handed end caps are colored red (blue). At zero temperature,
the two end caps then grow symmetrically until they meet near the
midpoint of the particle. While they grow, the area of large curl is
concentrated at the interface between the volume where the
magnetization has already aligned parallel with the applied field and
the volume where it remains antiparallel to the field.
Each interface is composed of two regions with opposite $z$-component 
of the curl, with a curling type of singularity in the center of the
pillar.  
These two regions are separated by a layer where the $z$-component of 
the curl is close to zero, and the magnetization vectors in the $x$$y$-plane have a negative divergence located at the center of the pillar.
 Because the
end caps have opposite curls, in the region where the end caps come into
contact the curl changes abruptly from large positive to large
negative values. Some time is required for this defect to
disappear. This reversal of pillars by nucleation of reversed volumes
at the ends followed by growth of the reversed regions is consistent
with minimization of the micromagnetic energy at zero temperature
\cite{YAN,BERTRAM} and Monte Carlo simulations at finite temperature,
\cite{CHUI} both under quasistatic field-sweep conditions.

\begin{figure}
\vskip 2.5in
\includegraphics{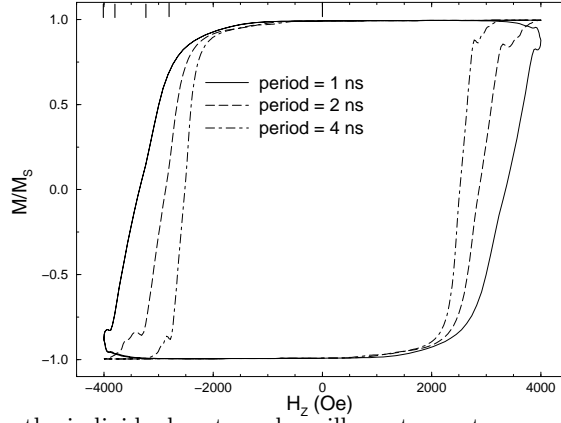}
\caption[] {Hysteresis loops for the individual rectangular pillars at
zero temperature and loop periods of $1$, $2$, and $4\,{\rm ns}$. The
change in the shape of the loops as the frequency is lowered indicates
that the magnetization is not following the applied field in a
quasi-static manner. The large tick marks on the upper horizontal
axis indicate the times for the images of the magnetization and its
curl, shown in Fig.~\protect{\ref{fig:hystpicture}}}
\label{fig:hystloop}
\end{figure}

The hysteresis loop associated with this simulation is shown in
Fig.~\ref{fig:hystloop}, with the field values shown in
Fig.~\ref{fig:hystpicture} indicated by large tick marks. Two periods are
shown for the $1\,{\rm ns}$ loop. Aside from differences due to the
initial alignment of the magnetization and the presence of the defect
after a complete loop, the reproducibility at zero temperature is
excellent.  Hysteresis loops under the same conditions, but for
periods of $2$ and $4$~ns, are also shown. These simulations show that
the hysteresis loop becomes more square for longer periods, indicating
that the magnetization is not following the applied field in a
quasi-static fashion. This is quite reasonable given that the
simulated loops correspond to microwave frequencies, but it makes the
infinite-period coercive field difficult to measure.

\begin{figure}
\vskip 2.5in
\includegraphics{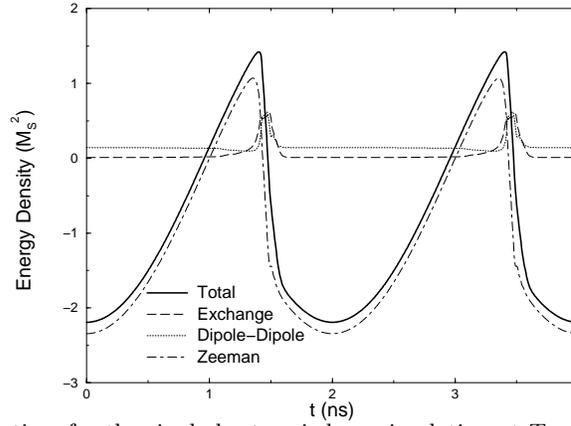}
\caption[] {Energy density vs time for the single hysteresis-loop
simulation at $T=0\,{\rm K}$ with period $4$~ns from
Fig.~\protect{\ref{fig:hystloop}}. Assuming the energy is near its
metastable minimum before the end caps start to propagate, the coercive
field can be estimated from the local maximum in the total energy.}
\label{fig:hystenergy} \end{figure}

The average energy density, along with its contributions, as functions
of time appear in Fig.~\ref{fig:hystenergy} for the $4$~ns hysteresis
loop. The energy density for zero crystalline anisotropy is calculated
as
\begin{equation}
\label{eq:energy}
E =
- \frac{1}{N} \sum_{i=1}^{N} 
    {\bf M}({\bf r}_{i}) \cdot 
    \left( \frac{1}{2}
       \left[ {\bf H}_{\rm e}({\bf r}_{i})
              +{\bf H}_{\rm d}({\bf r}_{i})
       \right]
      + {\bf H}_{\rm z}({\bf r}_{i})
    \right)
\;,
\end{equation}
where $N$ is the number of lattice sites. As the applied field begins
to decrease and then become negative, it is the Zeeman energy that
changes the most while the exchange and dipole-dipole energies remain
nearly constant. Near the coercive field, the exchange energy rapidly
increases as regions of reversed magnetization, and the interfaces
associated with them, form at the ends of the pillar. Once these
reversed regions become large enough, they grow spontaneously and the
Zeeman energy rapidly decreases, while the exchange energy from the
interface between reversed and unreversed regions remains roughly
constant. The growth of the reversed regions does not take long, and
the exchange energy quickly dissipates when the regions from the two
ends merge at the middle and the reversal is complete.

The information in Fig.~\ref{fig:hystenergy} can be used for
extracting the coercive field of the nanomagnet. For infinitely slow
variations in the applied field, ${d}E/{d}t$ is proportional to
${d}E/{d}{H}_{\rm a}$ using the chain rule. Thus portions of the
figure with positive slope of the total energy correspond to fields
for which the system is trapped behind an energy barrier, while
portions with negative slope correspond to fields for which the system
responds deterministically by finding a new energy-minimizing
magnetization configuration. Fields for which the energy has a maximum
should then correspond to the coercive field. Dynamic effects cause the
field at which this maximum occurs to depend on the frequency.

\begin{figure}
\vskip 2.5in
\includegraphics{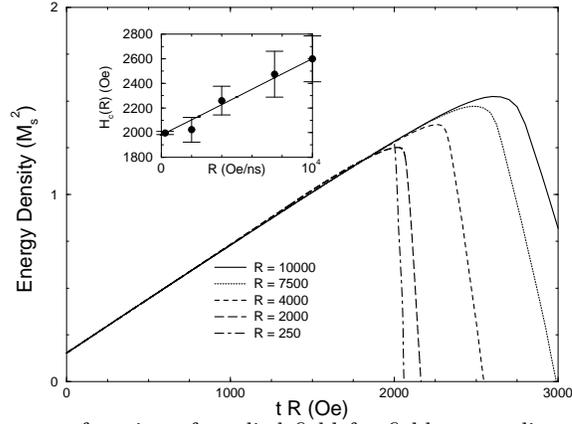}
\caption[] {Total energy density as a function of applied field for
fields swept linearly with rate $R$ at zero temperature. The estimation
of the zero-rate coercive field $H_c$ estimated from the energy maximum
is shown in the inset. The estimated value clearly depends on the rate
of change of the applied field, but using linear fitting the static
coercive field is found to be $1979 \pm 14 \,{\rm Oe}.$ The error bars
are estimated using the second derivative at the maximum.}
\label{fig:coercenergy} \end{figure}

To estimate the static coercive field we find
$H_c(R)$ for fields that vary linearly with time, $H(t)=-R t$ for
$t>0$, and then extrapolate to the $H_c(R=0)$ value. The energy
density is presented as a function of the absolute value of the
applied field in Fig.~\ref{fig:coercenergy} for rates ranging from
$R=250\,{\rm Oe/ns}$ to $R=10000\,{\rm Oe/ns}.$ The coercive fields
estimated from the energy maximum are shown in the inset where the
error bars are estimated from the second derivative near the energy
maximum. The coercive field clearly decreases with $R.$  
Extrapolation of a weighted least-squares fit yields a coercive field of
$H_c=1979 \pm 14 \,{\rm Oe}$ at $R=0.$ This is a more accurate
estimate of the static coercive field than those taken directly from
hysteresis-loop simulations.

\subsection{Switching in Constant Field}

\label{sec:pillarswitch}

\begin{figure}
\vskip 2.5in
\includegraphics{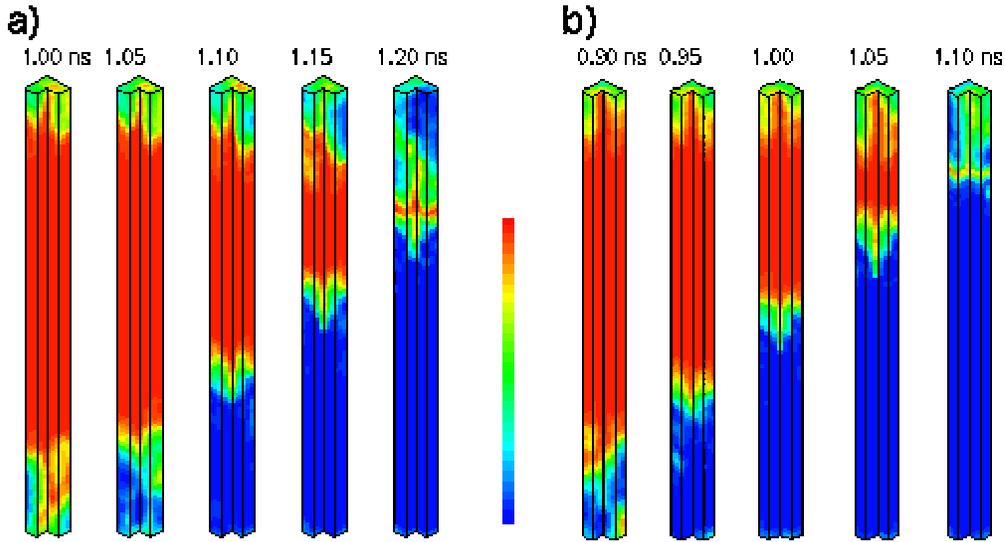}
\caption[] {
Magnetization reversal in the nanoscale magnet after a rapid reversal
of the field. The magnets have a square cross section, but are shown
in a one-quarter cut-away view. Both reversals are for $H=1800\,{\rm
Oe}$ and $T=20\,{\rm K}$. Here the $z$-component of the magnetization is
shown, with red representing the metastable orientation and blue
representing the equilibrium orientation.  (a) Nucleation of both end
caps, but at different times.  (b) Nucleation of one single end cap
that grows to reverse the entire magnet.  }
\label{fig:switchpicture}
\end{figure}

At finite temperature the magnetization can reverse, even when the
applied field is weaker than the coercive field. In this case, thermal
fluctuations carry the magnetization past the free-energy barrier and
on to the new equilibrium configuration. Two examples of such
thermally induced switching are shown in Fig.~\ref{fig:switchpicture}
for the conditions $H_0=1800$~Oe and $T=20$~K. For small positive
times (not shown) the volumes associated with the end caps fluctuate
due to the thermal noise. The reversal appears to proceed by a
nucleation process, with the end caps serving as the seeds for
heterogeneous nucleation. The initiation of the switching from the
end caps is similar to results for reversal induced by applied fields
greater than $H_c$ in experiments, \cite{KIRK} and simulations with
$T$$=$$0\,{\rm K}$ where reversal begins at the end caps \cite{BERTRAM,FANG}
or corners. \cite{DAO}

In classical droplet theory, in which boundaries such as the sides of
the magnet are ignored, nucleation of the equilibrium magnetization is
governed by the competition between the favorable Zeeman energy due to
alignment with the applied field and the unfavorable exchange energy
due to the interface with the majority volume of the misaligned
orientation. The exchange energy dominates for small droplets, which
tend to shrink. Large droplets tend to grow because the Zeeman energy
dominates. These two regimes are separated by a critical droplet size,
the saddle point associated with the free-energy barrier, where the
tendencies toward growth and shrinkage are balanced.
In nanoscale magnets the situation is more complicated because the
free energy cannot be easily separated into surface and volume
contributions. Nevertheless, the important aspect remains that a 
free-energy barrier exists that must be crossed for reversal to begin.
The end caps fluctuate until a succession of highly unlikely
fluctuations carries the end-cap configuration past the free-energy
barrier. After that, growth of the reversed region is energetically
favorable and occurs rapidly, on the order of $0.2$ to $0.3$~ns for the
pillars.

For long pillars, nucleation events at the two ends occur
approximately independently of each other. In addition, the growth of
a supercritical region takes a significant amount of time: enough time
for the other end cap to have a reasonable probability to nucleate and
begin to grow. One example of both ends nucleating at different times
is shown in Fig.~\ref{fig:switchpicture}(a), while
Fig.~\ref{fig:switchpicture}(b) shows nucleation of one end that grows
to switch the magnetization of the entire pillar. For the present
applied field the latter situation is somewhat rare, from the theory
presented below it can be inferred that it occurs in about 10\% of
simulated switches. From Fig.~\ref{fig:switchpicture}(b) it can also
be seen that the nucleation of the two end caps is not completely
independent. As one reversed region grows along the pillar, the
free-energy barrier at the other end cap decreases. Eventually the
free-energy barrier for the second end becomes zero and it will also
begin to grow. If a less negative field is applied, the free-energy
wells of the metastable states will be deeper and the first reversed
region will have to grow further along the pillar before this occurs.

\begin{figure}
\vskip 2.5in
\includegraphics{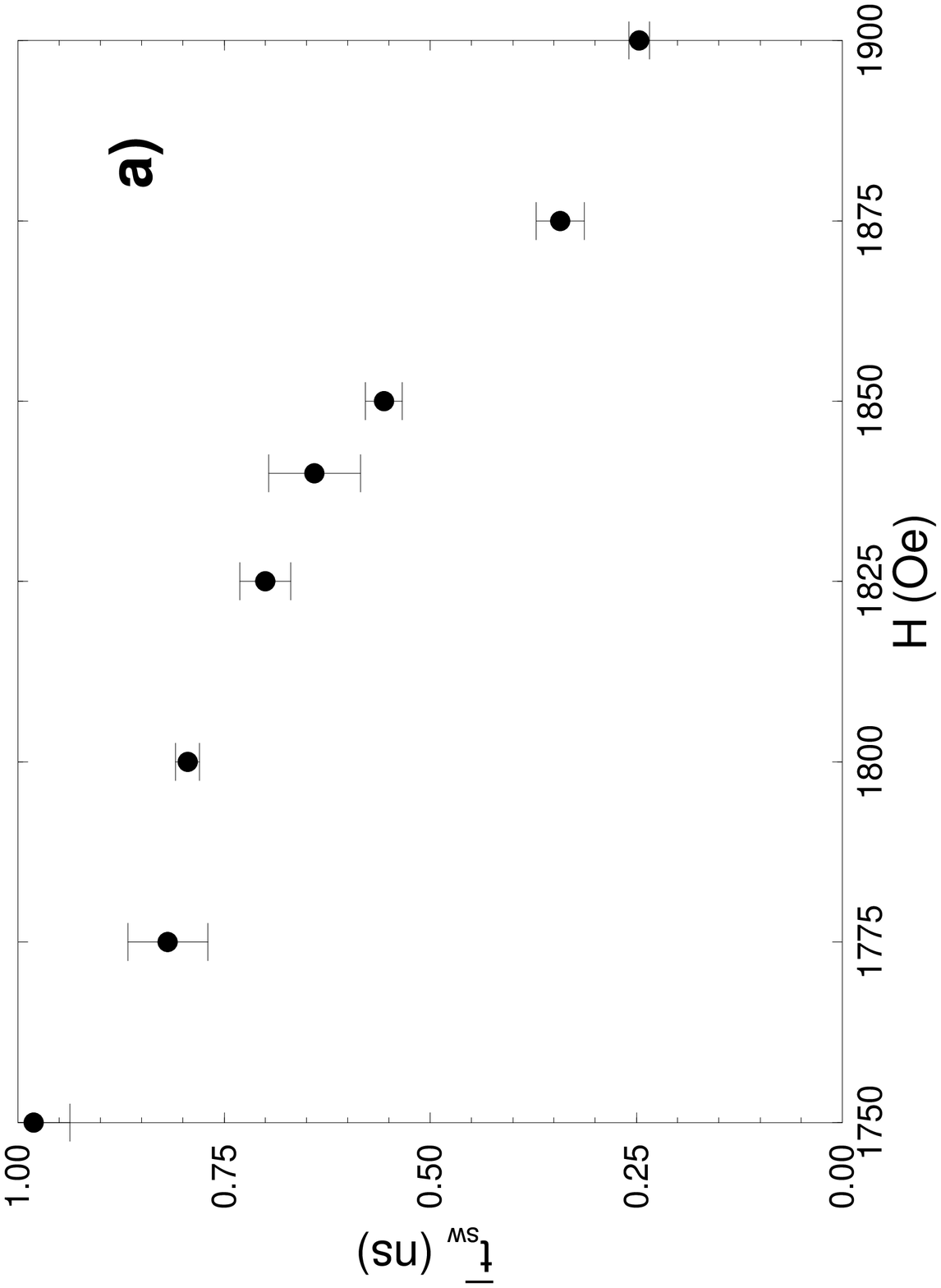}
\includegraphics{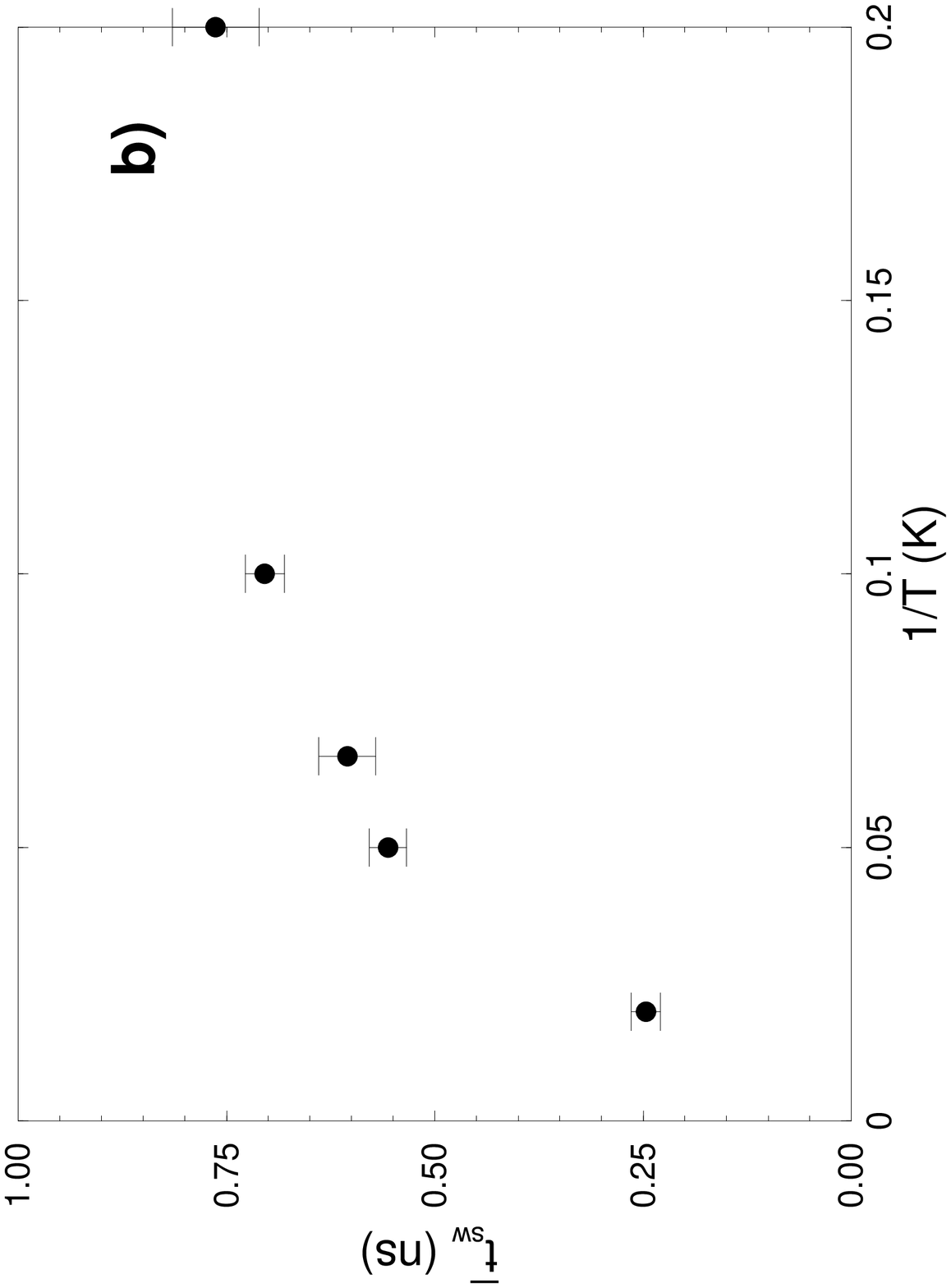}
\caption[] {Mean switching time as a function of (a) applied field at
$T=20\,{\rm K},$ and (b) temperature at $H=1850\,{\rm Oe}.$ The
switching time increases as (a) the free-energy barrier separating the
metastable and stable orientations grows as the applied
field is decreased, or (b) the thermal energy available for crossing
the barrier decreases as temperature is lowered.}
\label{fig:switchdependence}
\end{figure}

The mean switching time, $\bar{t}_{\rm sw}$, defined here as the first time
when $M_z = 0,$ is shown in Fig.~\ref{fig:switchdependence}(a) as a
function of applied field for $T=20\,{\rm K}.$ The difference in the
error bars reflects the amount of statistical sampling at the
different conditions. The dependence of $\bar{t}_{\rm sw}$ on temperature is
shown in Fig.~\ref{fig:switchdependence}(b) for applied fields of $H_0
= 1850\,{\rm Oe};$ fields are always applied along the pillar. While
there is a clear dependence on field and temperature, we find no clear
functional form for the dependence.  Notably, the exponential
dependence of $\bar{t}_{\rm sw}$ on inverse temperature, expected when the
barrier is high, is not seen.  This implies that the free-energy
barrier is low.  In fact, this picture is consistent with the
switching statistics described next, where the dynamics are consistent
with a biased walk in a shallow well.

\begin{figure}
\vskip 2.5in
\includegraphics{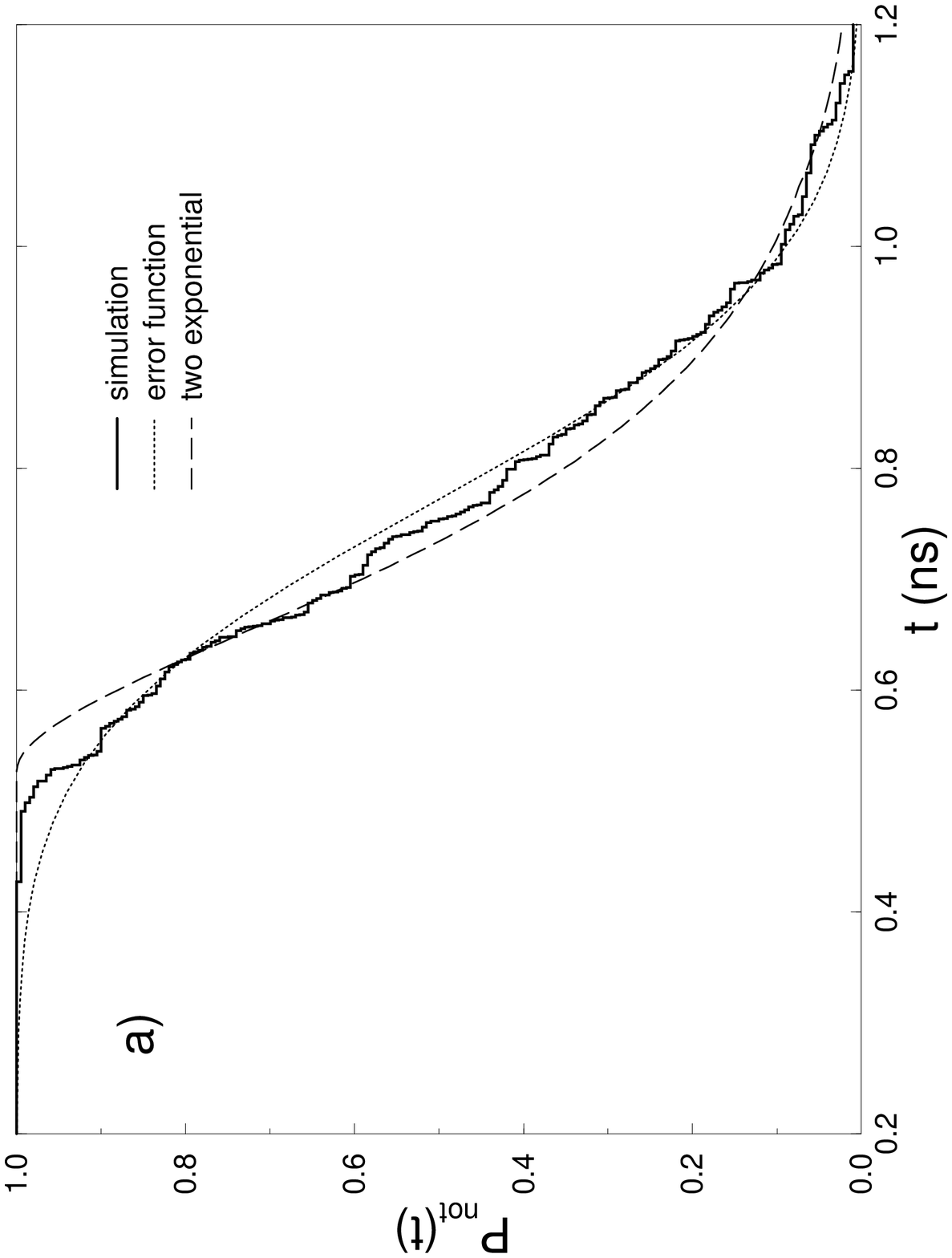}
\includegraphics{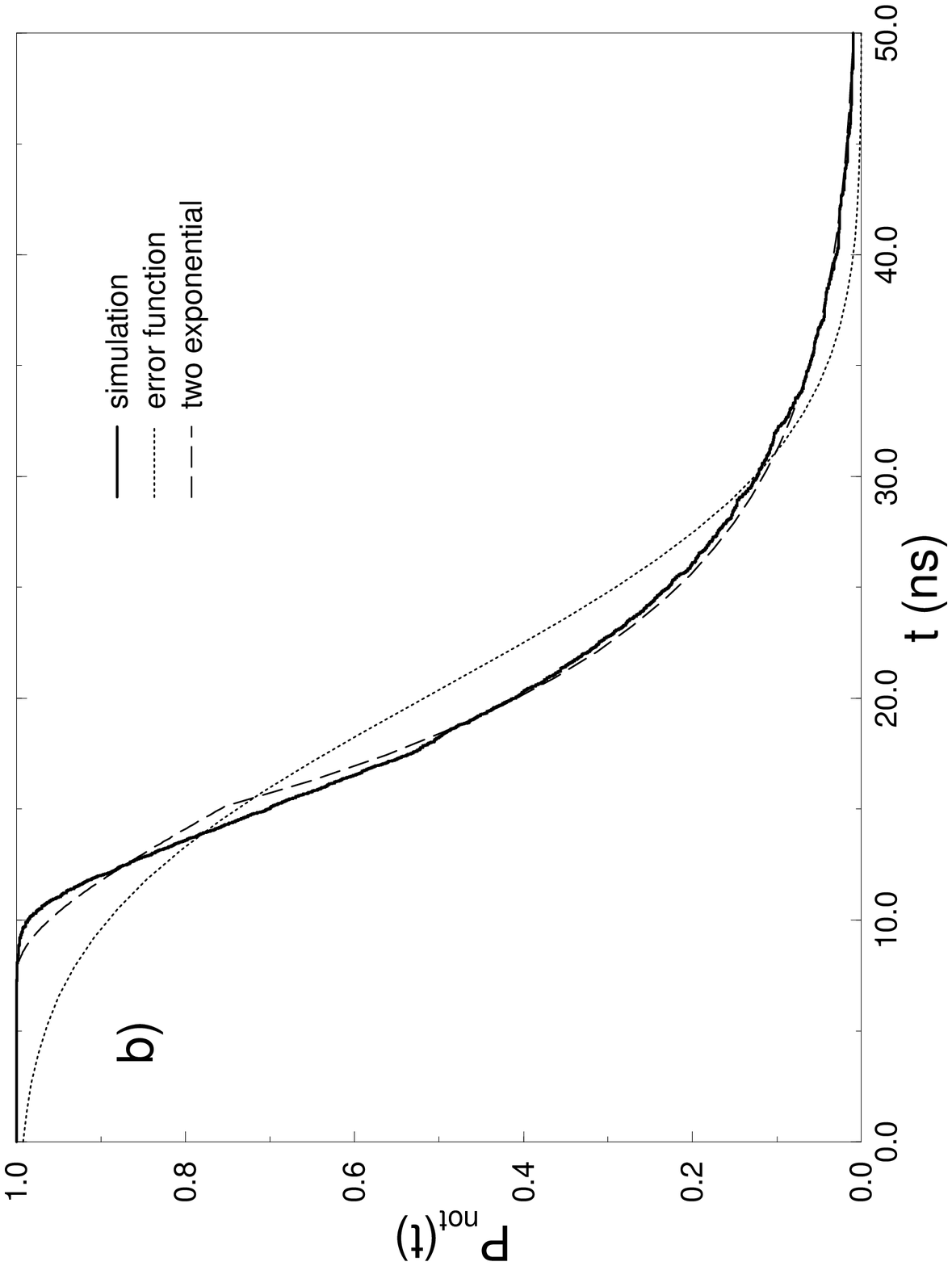}
\caption[] {Probability of not switching, $P_{\rm not}(t)$, for (a)
$200$ switches of nanoscale magnetic pillars at $H=1800$~Oe and
$T=20$~K, and (b) $2000$ switches in a simple one-dimensional model
nanopillars at $H=1000$~Oe and $T=20$~K. The theoretical forms,
Eqs.~(\ref{eq:twoexpo}) and (\ref{eq:Perf}), were fitted by matching
their first and second moments to those of the simulation data.}
\label{fig:pnot}
\end{figure}

The statistics of magnetization switching are measured by $P_{\rm
not}(t)$, the probability of not switching before time $t$.  The
simulation results for $P_{\rm not}(t)$ for $200$ switches at $H_0 =
1800 \,{\rm Oe}$ and $T=20\,{\rm K}$ are presented in
Fig.~\ref{fig:pnot}(a). The result is clearly not exponential, the
expected functional form when the free-energy barrier is
large. Experimental results for single-domain magnets, in which
$P_{\rm not}(t)$ is not exponential, have been reported recently.
\cite{KOCH}

\begin{figure}
\vskip 2.5in
\includegraphics{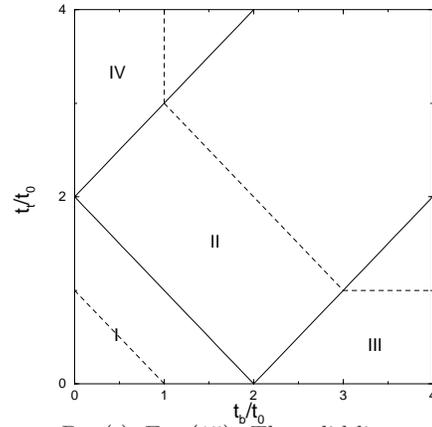}
\caption[] {Integration of weights to get $P_{\rm not}(t),$
Eq.~(\protect{\ref{eq:twoexpo}}). The solid lines separate regions where
one (or two) nucleation events occur before switching, while the dashed
curves are integration paths for constant switching time, $t_{\rm
sw}/t_0=3/2$ and $3$. See the text for a full explanation.}
\label{fig:integration} 
\end{figure}

A simple theory that assumes completely independent nucleation at the
two ends of the pillar produces a reasonable description of the simulation
result. The theory for $P_{\rm not}(t)$ requires two parameters. The
first is the constant nucleation rate, ${\cal I}$, for the formation
of a supercritical reversed region at one isolated end of a
pillar. The second is the rate, ${v}$, at which a single growing
region changes the normalized magnetization of the pillar. Labeling
the two independent nucleation times at the top and bottom $t_{\rm t}$
and $t_{\rm b}$, respectively, the positive values of the
magnetization of the pillar can be described by
\begin{equation} 
\label{eq:peom} 
M_z(t) = \left\{
\begin{array}{lr} 
1                       \qquad\qquad & t < t_1 \\ 
1-{v}(t-t_1)            \qquad\qquad & t_1 \le t < t_2 \\
1-{v}(t-t_1)-{v}(t-t_2) \qquad\qquad & t_2 \le t 
\end{array} 
\right.  
\;, 
\end{equation} 
where $t_1 = {\rm min}(t_{\rm t},t_{\rm b})$ and $t_2 = {\rm
max}(t_{\rm t},t_{\rm b})$. We define switching to occur at
$M_z(t_{\rm sw})=0$, when roughly one-half the volume of the
nanomagnet is oriented in the equilibrium direction. The switching
can occur either through one or both ends nucleating, and the
actual switching time for a particular $(t_{\rm t},t_{\rm b})$ is
given by
\begin{equation}
\label{eq:tswitch}
t_{\rm sw} = 
{\rm min}\left(2t_0+t_1,t_0+\frac{t_1+t_2}{2}\right)
\;,
\end{equation}
where $t_0=1/(2v)$ is the earliest time at which the pillar can switch.  For
times $t<t_0$ the probability of switching is zero, even if both ends
nucleate immediately after the reversal. The situation in the $t_{\rm
b}$-$t_{\rm t}$ plane is shown in Fig.~\ref{fig:integration}, where
the dashed curves correspond to two different conditions of constant
$t_{\rm sw},$ $t_{\rm sw}/t_0$$=$$3/2$ and $3.$ The solid lines
divide the plane into regions with different switching histories.  In
the triangle near the origin (I), switching occurs for times
$t_0<t_{\rm sw}<2t_0$, which can happen only if both ends nucleate.
Between the parallel solid lines (II) switching occurs by double
nucleation, while outside them (III \& IV) a single nucleation causes
the switching. Since each nucleation process has a constant rate, the
corresponding probability density function for the nucleation time is
exponential, ${\cal I}\exp{(-{\cal I}t)}$. Integrating along curves of
constant $t_{\rm sw}$, the probability of not switching before time
$t$ is found to be \cite{JAP2000}
\begin{equation}
\label{eq:twoexpo}
P_{\rm not}\left(t\right) = 
\left\{
\begin{array}{lr}
  1                                 \qquad\qquad & t< t_0 \\
  e^{-2{\cal I}(t-t_0)} 	    
    \left[1+2{\cal I}(t-t_0)\right] \qquad\qquad & t_0 \le t< 2t_0 \\
  e^{-2{\cal I}(t-t_0)} 	    
    \left[1+2{\cal I}t_0\right]     \qquad\qquad & {2t_0\le t}
\end{array}
\right.
\;.
\end{equation}
We will refer to this as the ``two-exponential'' decay model.
In the limit of infinitely fast growth, $t_0=0,$ only regions III and
IV remain, and this becomes a simple exponential decay associated with
thermally activated crossing of a free-energy barrier. For $t_0>0$,
$P_{\rm not}(t)$ is quadratic in $t-t_0$ for $t \rightarrow t_0^+$,
has a discontinuity in the slope at $t=2t_0$, and is exponential for
$t>2t_0$.

The fit of the two-exponential decay model of Eq.~(\ref{eq:twoexpo})
to the simulation results appears in Fig.~\ref{fig:pnot}(a) as the dashed
curve. The dotted curve is a fit to the error function
\begin{equation}
\label{eq:Perf}
P_{\rm erf}(t) = 
\frac{1}{2} - \frac{1}{2}{\rm Erf}\left(\tilde{\cal{I}}(t-\tilde{t}_0)\right)
\;,
\end{equation}
which was chosen as an empirical form because it also has two
adjustable parameters, chosen here to have a similar interpretation as
the theory given above. The parameters for both theoretical forms were
determined by matching the first and second moments of the theoretical
form to the moments of the simulation data, and the parameters
are
${\cal{I}}       \approx 4.049$, 
$t_0             \approx 0.527$ and 
$\tilde{\cal{I}} \approx 4.176$,
$\tilde{t}_0     \approx 0.790$.
For these conditions, both forms, Eqs.~(\ref{eq:twoexpo}) and
(\ref{eq:Perf}), seem to fit the simulation data about equally
well. The disagreement with Eq.~(\ref{eq:twoexpo}) probably occurs
because the condition that the free-energy barrier is much greater
than $k_{\rm B}T$ is not met. When the free-energy barrier is much
less than $k_{\rm B}T,$ the magnetization should be more like a biased
random walk. In that case the most noticeable effect of thermal
fluctuations would be a symmetric smearing of the switching times
around $\bar{t}_{\rm sw}.$

The question of statistical sampling is also important. Specifically,
one expects to have undersampling of the population because the
finite-sampling $P_{\rm not}(t)$ goes to zero faster than the
infinite-sampling $P_{\rm not}(t)$ with increasing $t$.  In fact, for
times less than the maximum observed switching time in the long-time
tail of the distribution, the error function underestimates the
population, while the two-exponential theory overestimates it. To this
extent, it is possible that data with better sampling may shift
towards better agreement with Eq.~(\ref{eq:twoexpo}). Indeed,
experimental switching probabilities \cite{KOCH} in single-domain
magnets have been observed to have exponential tails, which is
inconsistent with Eq.~(\ref{eq:Perf}).

Better statistics and a higher free-energy barrier are needed to
validate the theoretical $P_{\rm not}(t)$ given in
Eq.~(\ref{eq:twoexpo}).  For the present model, better statistics would
be prohibitive; the $200$ switches considered here took approximately
$20$ weeks to generate using $4$ processors on an Origin 2000. A less
realistic but numerically faster simulation, in which the pillars are
modeled as one-dimensional stacks of cubes, is presented in the next
section. The advantage of this simplified model is that it allows many
more switching events to be observed in a reasonable amount of
computer time, even with much higher free-energy barriers. This improves
the statistics for the very rare events with $t_{\rm sw} \gg 2t_0$ and
$t_{\rm sw} \rightarrow t_0$, which are important to distinguish
between alternative theoretical forms for $P_{\rm not}(t)$.

\section{A Simple Model}

\label{sec:1dpillar}

In order to generate a sufficiently large sample of switching events
in nanoscale magnets, we consider a simpler model in which the pillar
consists of a one-dimensional stack of cubic cells, which is the model
introduced by Boerner and Bertram. \cite{BandB} The cubes have a
linear size $\Delta \tilde{r} = 2 \ell_{\rm e},$ here $5.2\,{\rm nm},$
since the material parameters for iron are used (Ref.~\onlinecite{BandB} used
parameters for nickel). To keep the aspect ratio of the nanopillars as
close as possible to those considered above, a pillar composed of $17$
cubes is used. For this simple model, the local field due to the
dipole interactions is calculated as \cite{BandB}
\begin{equation}
\label{eq:bbdipole}
{\bf \tilde{H}}'_{\rm d}({\bf \tilde{r}}_i) =
(\Delta \tilde{r})^3 
\sum_{j \ne i} 
\frac{3{\bf\hat{r}}_{ij}
     \left(
     {\bf\hat{r}}_{ij} \cdot {\bf\tilde{M}}({\bf\tilde{r}}_j)
     \right)
     -{\bf\tilde{M}}({\bf\tilde{r}}_j)}
   {{\bf\tilde{r}}_{ij}^3}
\;,
\end{equation}
where ${\bf\tilde{r}}_{ij}$ is the displacement vector from the center
of cube $i$ to the center of cube $j,$ and ${\bf\hat{r}}_{ij}$ is the
corresponding unit vector. The factor of volume $(\Delta \tilde{r})^3$
results from integrating over the constant magnetization density in
each cell. This factor can be combined with $\tilde{r}_{ij}^3$, so
that the denominator depends only on
${\bf{x}}_{ij}={\bf{\tilde{r}}}_{ij}/(\Delta \tilde{r})$, which is a
vector of integers denoting the difference in cell indices.  It can
readily be seen that for a uniformly magnetized pillar the summation
over index integers leads to different values of ${\bf H}'_{\rm
d}({\bf r})$ for different choices of discretization length along the
long axis.  This happens because self-contributions have been ignored
and the far-field result has been used for neighboring cells. However,
since this model is being used to investigate only the statistics of
switching, and not for estimating physical values, the results should
be qualitatively correct.  Using the scaling to dimensionless units,
the approximate dipole field is
\begin{equation}
\label{eq:rbbdipole}
{\bf {H}}'_{\rm d}({\bf{r}}_i) =
\sum_{j \ne i} 
\frac{3{\bf\hat{r}}_{ij}
  \left(
    {\bf\hat{r}}_{ij}\cdot{\bf{M}}({\bf{r}}_j)
  \right)
  -{\bf{M}}({\bf{r}}_j)}
  {{\bf{x}}_{ij}^3}
\;.
\end{equation}
We have verified that our implementation of the model agrees with that
of Boerner and Bertram by reproducing the observed coercive field for
the nickel pillars discussed in Ref.~\onlinecite{BandB}. Because of
the different approximations, the coercive field for the present iron
pillars is about $1500\,{\rm Oe}.$

The probability of not switching for $2000$ switches for this model of
iron pillars is shown in Fig.~\ref{fig:pnot}(b), along with the two
theoretical forms, Eqs.~(\ref{eq:twoexpo}) and (\ref{eq:Perf}), which
have been fit using the same procedure as in
Sec.~\ref{sec:pillar}. The kink in the two-exponential theoretical
form at $2t_0$ is quite noticeable, but cannot be seen in the
simulation data. The kink has its origin in the complete suppression
of nucleation for negatives times, and the absence of the kink in the
simulation data may stem from the finite chance of nucleation at
negative times due to the way the field is reoriented, as described in
Sec.~\ref{sec:pillarcoerc}. Another possibility is that the
interactions between the ends smooth out the difference between the
one- and two-nucleation decay modes. Despite the kink, the theoretical
form Eq.~(\ref{eq:twoexpo}) does a good job of describing both the
exponential tail and the rounding at early times of $P_{\rm not}(t)$.
It is clearly superior to the error-function form, as well as to a
displaced, single exponential (not shown).

\section{Arrays of nanomagnets}

\label{sec:array}

The long-ranged dipole-dipole interactions that contribute to the
shape anisotropy of nanoscale magnets also cause interactions between
nanomagnets. This is especially true in most potential applications,
where miniaturization will drive devices to high densities. In
addition, the interactions between nanomagnets in arrays could be the
basis of device applications, prototypes of which have already been
investigated. \cite{DUNIN,SCIENCE}  Regular arrays of nanomagnets have
already been used experimentally to provide magnetic signals strong
enough to be measured \cite{WIRTH98,WIRTH99} in the experiments our
nanomagnets are modeled after. Our simulations show that even for very
wide spacings the magnetic interactions between nanomagnets have
significant effects on the switching properties.

\begin{figure}
\vskip 2.5in
\includegraphics{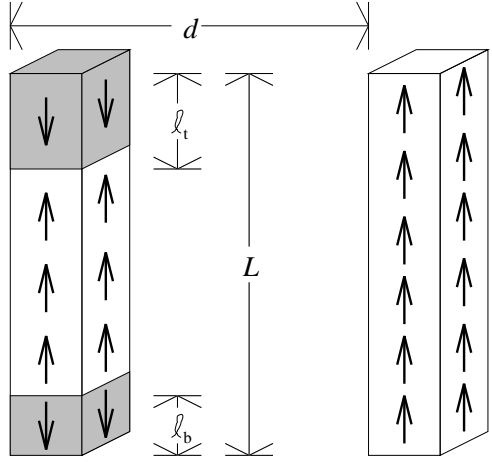}
\caption[] {Schematic of two pillars, one with a growing region of
magnetization in the equilibrium orientation at each end. The distance
between pillars is $d$, their height is $L$, and the length of the
reversed regions along the long axis are $l_t$ and $l_b$,
respectively.}
\label{fig:pillarscheme}
\end{figure}

To allow for analytic treatment, we consider only the two leading
contributions to the local dipole field ${\bf H}_{\rm d}({\bf
r})$ from the dipole and quadrupole moments of the source magnet. The
former contributes uniformly throughout the observation volume, while
the latter changes linearly in each Cartesian direction. Specifying
the distance between pillars as $d$ and considering the dipolar and
quadrupolar moments of the source magnet, $M_1^0$ and $M_2^0$,
respectively, the leading contributions to the observed demagnetizing
field are ${\bf H}_{\rm d}^{(1)}({\bf r}) = - {\bf\hat e}_{\rm z}
M_1^0 / d^3$ and ${\bf H}_{\rm d}^{(2)}({\bf r}) = - 9 (2z{\bf\hat
e}_{\rm z} - x{\bf\hat e}_{\rm x} - y {\bf\hat e}_{\rm y}) M_2^0 /(4
d^5),$ respectively. Simple expressions for the multipole moments can
be calculated in the following way. Assume a pillar with regions of
uniform magnetization oriented in the $+{\bf\hat e}$ direction in the
middle and in the $-{\bf\hat e}$ direction at the ends, as shown in
Fig.~\ref{fig:pillarscheme}. With the total pillar length $L$, the
top-region length $\ell_{\rm t}$, and the bottom-region length
$\ell_{\rm b}$ all measured along the long axis of the pillar, the
dipole moment for a reversing pillar is
\begin{equation}
\label{eq:m1}
M_1^0 = M_{\rm S} A 
  \left( L - 2 {\ell}_{\rm t} - 2{\ell}_{\rm b}
  \right)
\end{equation} 
and the quadrupole moment is
\begin{equation}
\label{eq:m2}
M_2^0 = 2 M_{\rm S} A 
   \left(\ell_{\rm t}^2 - \ell_{\rm b}^2 
         - L \left( \ell_{\rm t} - \ell_{\rm b}
             \right)
   \right)
\;,
\end{equation}
where $A$ is the cross-sectional area of the pillar. The dimensionless
form is achieved by setting $M_{\rm S}$ to unity and using
dimensionless lengths.

Using the simplified expression for the dipole moment in
Eq.~(\ref{eq:m1}), the contribution from one upward-magnetized pillar
at the nearest-neighbor position in the array is ${\bf H}_{\rm
d}^{(1)} \cdot {\bf\hat e}_{\rm z} \approx -1$~Oe. The quadrupole
contribution from Eq.~(\ref{eq:m2}) at the end of one nanomagnet that
is the nearest neighbor to a switching pillar with $\ell_{\rm t}=L/2$
is ${\bf H}_{\rm d}^{(2)} \cdot {\bf\hat e}_{\rm z} \approx
0.4$~Oe. This is a rough estimate of the maximum interaction through
the quadrupole moment, and the sign indicates a tendency for
neighboring pillars in the array to switch at opposite ends.

As a simple initial investigation, we consider a linear array of four
of the rectangular nanomagnets described in Sec.~\ref{sec:pillar},
with the nanomagnets oriented perpendicular to the substrate. Since
dipole-dipole interactions within each nanomagnet are calculated using
the Fast Multipole Method described in Appendix~1, the multipole
moments for each nanomagnet are readily available. These moments can
be used to quickly calculate the interactions between nanomagnets in
the array, under the constraint that the far-field description is
appropriate. To ensure this we have considered only a spacing between
pillars of two pillar lengths, $300\,{\rm nm}.$ Note that this
situation is quite different from Fast Fourier Transform approaches,
in which the calculation must be carried out on a lattice that also
fills the space {\em between} the nanomagnets. \cite{RAMSTOCK} To do
that practically, the space between the magnets must be kept small.

To study arrays of nanomagnets, systems consisting of four $9$~{\rm
nm}~$\times$~$9$~{\rm nm}~$\times$~$150$~{\rm nm} parallel pillars
arranged in a line perpendicular to their long axes and spaced
$300$~nm apart were simulated using the Fast Multipole Method
truncated at $p=3.$ Hysteresis loops with periods on the order of a
few nanoseconds for the individual pillars in the array look similar
to those for isolated pillars in Fig.~\ref{fig:hystloop}, with no
observable difference between pillars on the outside and those on the
inside of the array. The symmetry equivalence for the two pillars on
the inside, and for the two pillars on the outside, will be used
throughout to double the statistical sampling.

\begin{figure}
\vskip 2.5in
\includegraphics{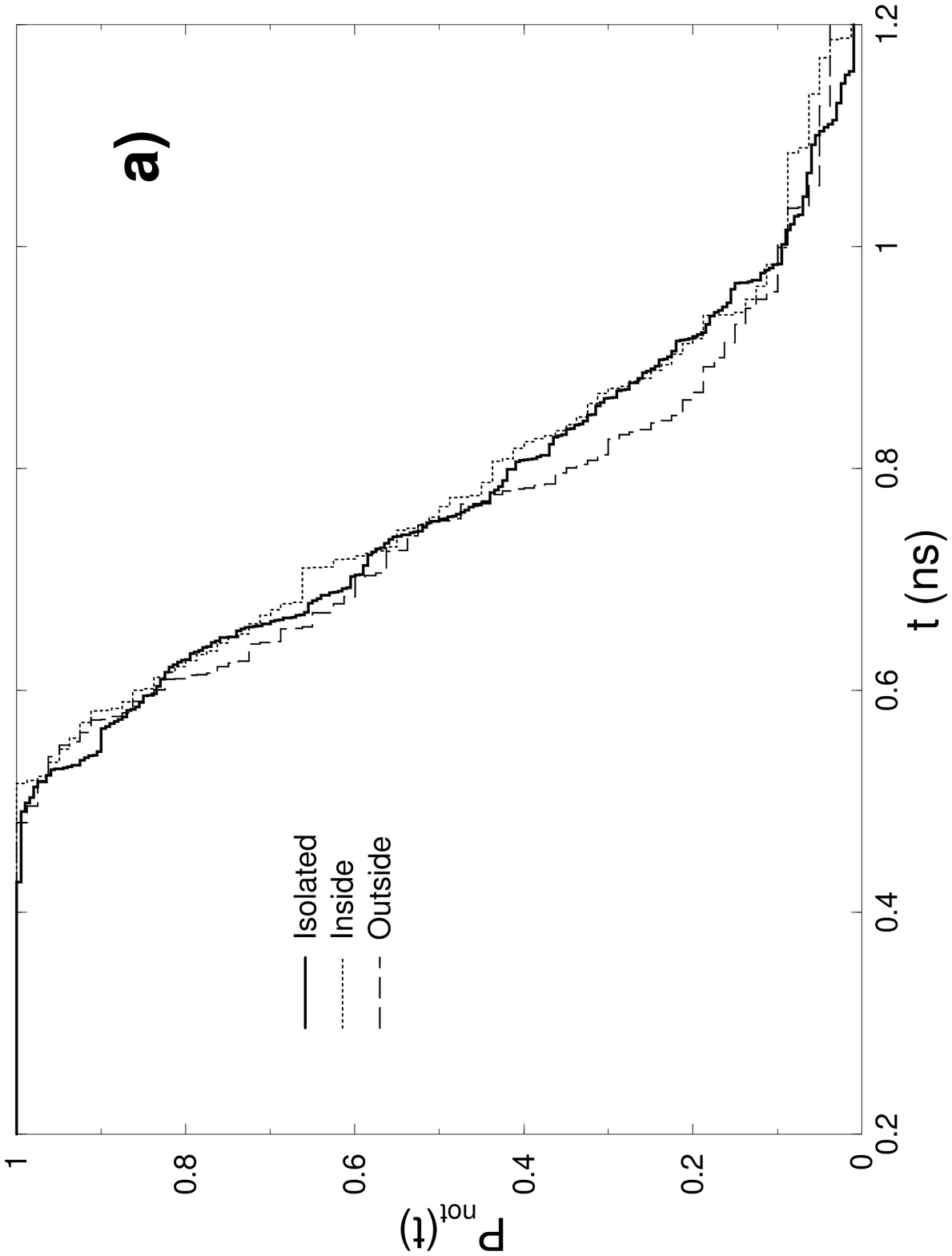}
\includegraphics{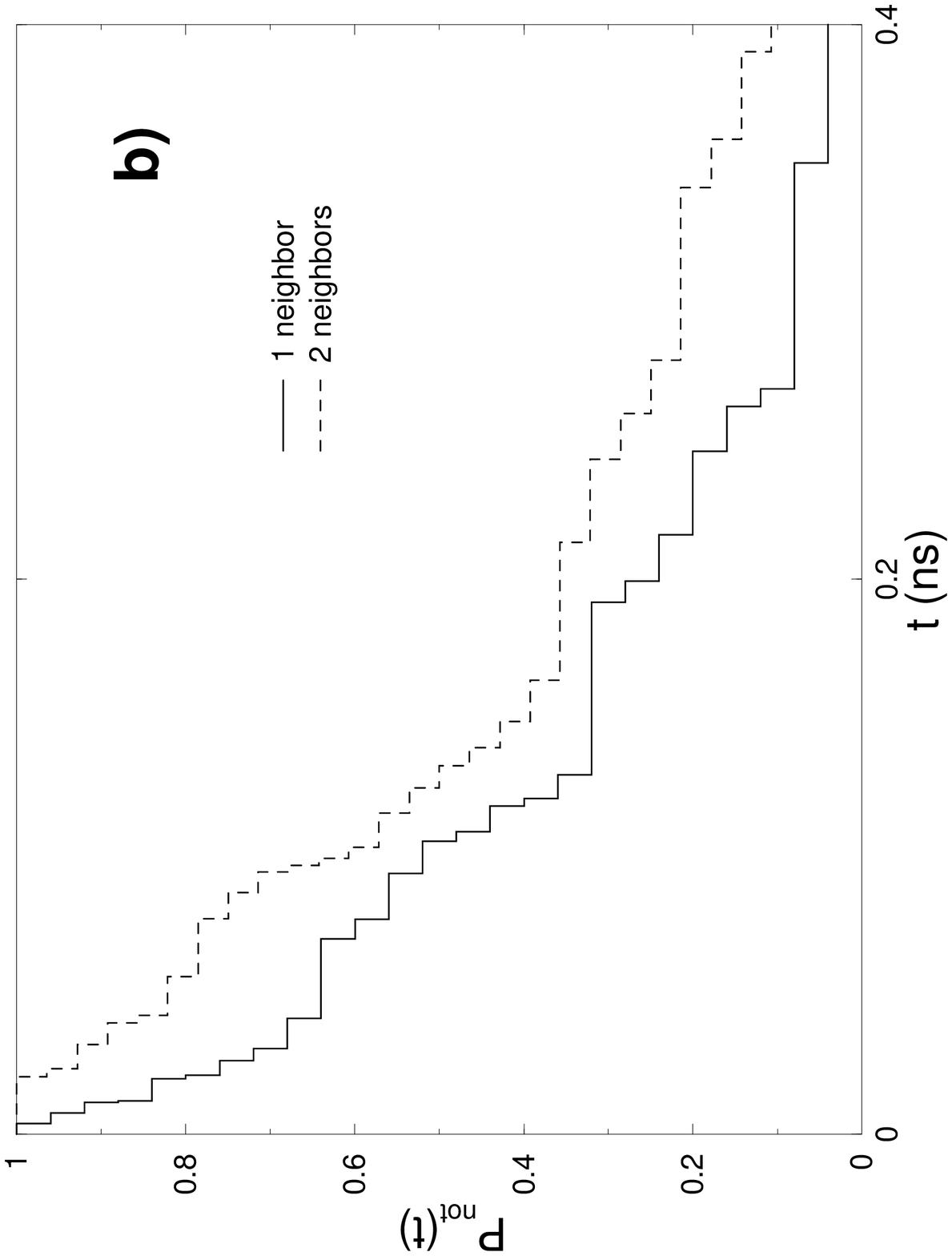}
\caption[] {(a) Probability of not switching, $P_{\rm not}(t)$, for
nanoscale magnetic pillars in a four-pillar array at $H=1800$~Oe and
$T=20$~K. The two pillars on the outside, and the two on the inside,
are equivalent by symmetry. The isolated pillar data are the same as
in Fig.~\protect{\ref{fig:pnot}}. (b) Interactions between pillars as
seen in the difference between $P_{\rm not}(t)$ for inside pillars
with one and both neighboring pillars already switched. Here $t$$=$$0$
corresponds to the time of the most recent switch of a neighboring
pillar.}
\label{fig:pnot2}
\end{figure}

The probability of not switching for $H=1800$~Oe and $T=20$~K is shown
for $40$ array switches in Fig.~\ref{fig:pnot2}(a) for pillars on the
inside and outside of the array, as well as isolated pillars.  No
significant difference can be seen between the three curves.  However,
the coupling between the pillars can be seen in the difference between
the $P_{\rm not}(t)$ for inside pillars with one or both nearest
neighbor pillars switched, shown in Fig.~\ref{fig:pnot2}(b).  Here,
$t$ is the time difference between $t_{\rm sw}$ and the last time a
neighboring pillar switched.  From this data it can be seen that of
pillars with two neighbors, those with only one neighbor switched tend
to switch earlier than those with both pillars switched. The effect is
even more pronounced in simulations of the simple model at
$H$$=$$1000\,{\rm Oe}.$ \cite{MMM01}

\section{Summary}

\label{sec:summary}

Numerical simulation of the Landau-Lifshitz-Gilbert micromagnetic
model has been used to investigate spatially non-uniform magnetization
switching in nanoscale magnets at the nanosecond time scale.  We have
focused on iron pillars $9 \, {\rm nm} \times 9 \, {\rm nm} \times 150
\, {\rm nm}$ because such pillars and arrays have been constructed and
measured experimentally. \cite{WIRTH98,WIRTH99} The zero-temperature
static coercive field has been estimated numerically to be $H_c=1979
\pm 14 \, {\rm Oe}$ by finding the field of maximum energy for fields
swept at constant rate, and then extrapolating to find the zero-rate
estimate.  Simulations of thermally activated magnetization switching
are possible at fields well below the coercive value. For the pillars
studied here, reversal occurs through nucleation at the ends of the
pillars. The probability of not switching, $P_{\rm not}(t),$ is not
well described by a delayed exponential, and a theoretical form,
Eq.~(\ref{eq:twoexpo}), based on independent nucleation at the ends of
the magnet is developed here. The agreement with results for
intensive, fully three-dimensional simulations with an applied field
near the coercive value are reasonable, but an {\em ad hoc} error
function gives similar agreement.  The agreement with
Eq.~(\ref{eq:twoexpo}) is much better when the field is well below the
coercive value and the statistics are better, which currently we have
only studied with less-intensive simulations that discretize the
nanomagnet only along its long axis.

The fully three-dimensional micromagnetics program has been developed
for massively parallel computers and implemented using the Fast
Multipole Method. These two features make it feasible to simulate
nanomagnets in widely spaced arrays. Even though the interactions
between the magnets are quite weak for the specific linear array
considered here, there are significant effects on the statistics of
the magnetization switching. Specifically, there is a dependence of
$P_{\rm not}(t)$ for a given pillar on the orientation of the
magnetization of its nearest neighbors in the array.  The nature of the
cooperative reversal mode observed here is a topic for future
research.

\section*{Acknowledgments}

This work was supported by NSF Grant No. DMR-9871455, Florida State
University (FSU) Center for Materials Research and Technology, the FSU
Supercomputer Computations Research Institute (U.S. DOE contract
number DE-FC05-85ER25000), and the FSU School of Computational Science and
Information Technology. Extensive supercomputer resources were
provided by FSU Academic Computing and Network Services and by the
U.S. Department of Energy through the U.S. National Energy Research
Scientific Computing Center.

\section*{Appendix 1}

\begin{figure}
\vskip 2.5in
\includegraphics{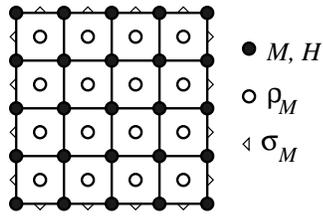}
\caption[] {Schematic of the relationship between the lattice and the
dual lattices, projected along the $z$-axis. The magnetization density
${\bf M}({\bf r}_i)$ and the local magnetic field ${\bf H}({\bf r}_i)$
are known at the lattice sites of the simple cubic lattice (solid
circles), the magnetic charge density $\rho_{\rm M}$ is known at the
dual lattice sites (open circles) which are located at the body center
positions, and the magnetic surface charge density is known at the
centers of squares defined by the surface lattice sites (open
triangles).} 
\label{fig:dual}
\end{figure}

Calculating the dipole-dipole interactions is the most intensive part
of the numerical calculation. The magnetic potential approach used
here involves defining a magnetic charge density $\rho_{\rm M}({\bf
r})=-{\bf \nabla} \cdot {\bf M}({\bf r}).$ (We present this only in
terms of the dimensionless quantities.) This charge was evaluated on a
cubic lattice dual to that of ${\bf M}({\bf r}_i),$ see
Fig.~\ref{fig:dual}, using equally weighted two-point differences,
specifically
\begin{equation}
\label{eq:rho}
\rho_{\rm M}({\bf {r_{\rm d}}}) = 
  -\sum_{\mu=1}^{3} \left(
  \frac{1}{4\Delta {r}}
  \sum_{|{\bf h|}=\sqrt{3}\Delta r/2}
  {\rm sign}\left({\bf h} \cdot {\bf\hat{e}}_{\mu} \right)
  {\bf M}\left({\bf r}_d + {\bf h}\right) 
  \cdot {\bf\hat{e}}_{\mu}  \right)
\;,
\end{equation}
where ${\bf r }_{\rm d}$ is a position on the dual lattice, $\sum_{\bf
h}$ is the sum over the corresponding corners of the cube from the
direct lattice, and ${\bf\hat{e}}_{\mu}$ are the Cartesian unit
vectors. If we were to apply Eq.~(\ref{eq:rho}) with ${\bf r}_{\rm d}$ just
outside the magnetic material, it would give a nonzero charge there.
To avoid this unphysical result, we have moved this charge to the
surface by defining a surface magnetic charge density $\sigma_{\rm
M}({\bf r}) = {\bf\hat{s}}({\bf r}) \cdot {\bf M}({\bf r}),$ where
${\bf\hat{s}}$ is the unit vector directed out of the surface. We have
considered surface charges only on the surface of the model magnet,
and we evaluate them at the centers of the squares defined by adjacent
points on the surface of the direct lattice. The four corners are
equally weighted so that
\begin{equation}
\label{eq:sigma}
\sigma_{\rm M}({\bf r}_{\rm d'}) = 
  \frac{1}{4} \sum_{|{\bf h}|=\sqrt{2}\Delta r/2} {\bf\hat{s}} \cdot
{\bf M}({\bf r}_{\rm d'}+{\bf h})
\;,
\end{equation}
where $\sum_{\bf h}$ runs over the four corners.  The numerical
approach of Eqs.~(\ref{eq:rho}) and (\ref{eq:sigma}) ensures that
there is no net magnetic charge on the system as a whole.

The magnetic potential $\phi_{\rm M}({\bf r})$ is found by integrating
over both the volume and surface charges, \cite{JACKSON}
\begin{equation}
\label{eq:potint}
\phi_{\rm M}({\bf r}) = 
\int_{\rm V} {d}{\bf r}' 
   \frac{\rho_{\rm M}({\bf r}')}{|{\bf r}-{\bf r}'|}
+\oint_{\rm S} {d}{\bf\hat{s}'} 
   \frac{\sigma_{\rm M}({\bf r}')}{|{\bf r}-{\bf r}'|}
\;.
\end{equation}
Numerically, such an operation can be quite expensive since
unsophisticated algorithms will require $O(N^2)$ operations, where $N$
is the number of lattice sites.  An algorithm that remains reasonable
for large systems is the Fast Multipole Method. \cite{GREENGARD}

We have chosen to calculate the magnetic potential using the Fast
Multipole Method (FMM) because it has several advantages over the more
traditional Fast Fourier Transform (FFT) approach. The biggest
difference between the two methods is that the FMM makes no
assumptions about the underlying lattice, while the FFT method assumes
a cubic lattice with periodic boundary conditions. One consequence of
this assumption is that numerical models of systems without periodic
boundary conditions require empty space around the magnet so that the
boundary conditions do not affect the calculation. The FFT also
requires the lattice to continue into regions of empty space that lie
between elements of an array of magnets. By contrast, a FMM
implementation only needs to consider volumes occupied by magnetic
material. It does not need any padding.  In addition to these
advantages, the FMM is more efficient for large numbers of lattice
sites.

The popularity of the FFT approach stems from the fact that it takes
$O(N\ln{N})$ calculations to evaluate Eq.~(\ref{eq:potint}). The FMM
has a larger overhead, but requires only $O(N)$ operations to
calculate the same potential. \cite{GREENGARD} This means an FFT
approach to Eq.~(\ref{eq:potint}) makes sense for small, cubic
lattices, but that the FFM approach will be more efficient for large,
irregular, or incomplete lattices.

The FMM algorithm exploits the fact that $\phi_{\rm M}$ at each
lattice point can be expanded in terms of spherical harmonics,
\begin{equation}
\phi_{\rm M}(r,\theta,\varphi) = \sum_{i=0}^{\infty} \sum_{j=-i}^{i}
  \left( L_i^j r^i + \frac{M_i^j}{r^{i+1}} \right)
  Y_i^j\left(\theta,\varphi\right)
\;,
\end{equation}
where the $L_i^j$ terms can be used to represent the potential close
to the lattice point, and the $M_i^j$ terms can be used to represent
the potential far from it, but not both simultaneously. In this
context, near ${\bf r}_i$ means being closer to ${\bf r}_i$ than any
other lattice point, and far means distances more than twice the
largest distance from the center of cell $j$ to any of its boundary
points. Following Greengard, \cite{GREENGARD} we define the spherical
harmonics
\begin{equation}
\label{eq:ylm}
Y_i^j(\theta,\varphi) =
\sqrt{\frac{(i-|j|)!}{(i+|j|)!}}
P_i^{|j|}(\cos{\theta})
e^{Ij\varphi}
\;,
\end{equation}
with $P_i^{j}(x)$ the associated Legendre polynomial, and
$I=\sqrt{-1}$.  Actually implementing this approach requires a
truncation of the expansion in $i$ at order $p$. We have found that the
demagnetizing field for $p=3$ is within $1\%$ of the exact value for
our simulations.

\begin{figure}
\vskip 3.5in
\includegraphics{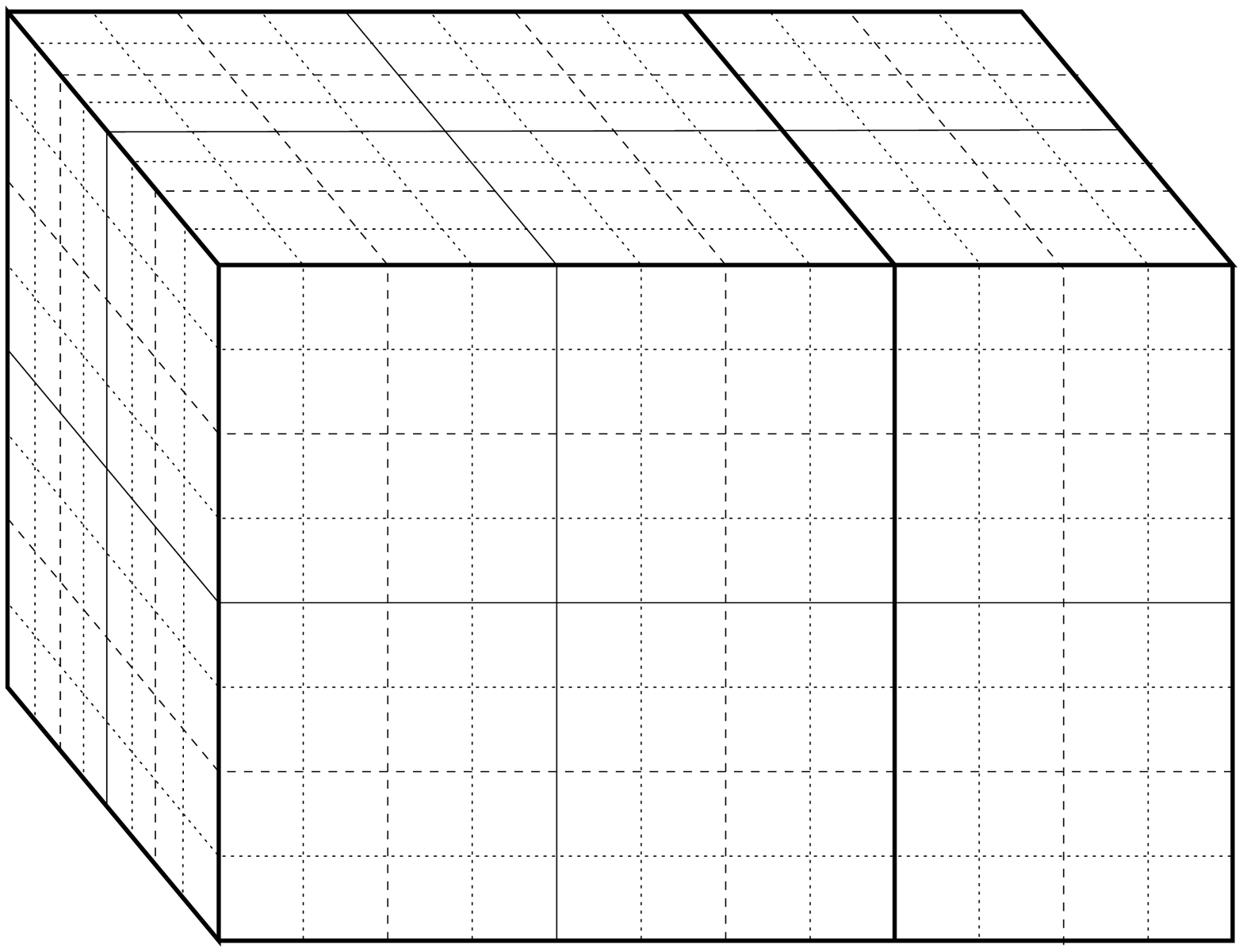}
\caption[] {Schematic of the hierarchical decomposition of space
chosen for our implementation of the Fast Multipole Method. This
eight-fold decomposition at each level is quite efficient given the
underlying cubic lattice.}
\label{fig:hierarchy}
\end{figure}

Our implementation of the FMM algorithm starts by partitioning the
model space into a system of cubes, an example is shown in
Fig.~\ref{fig:hierarchy}. The length of the side of the smallest
cubes, which are separated by dotted lines, is the same as the
discretization length. For our lattice system we have found that the
most efficient choice is to use cubes centered on the {\em direct}
lattice. For each cube, the multipole expansion coefficients of the
far-field potential are calculated for the specific configuration of
magnetic charges $\rho_{\rm M}$ and $\sigma_{\rm M}$
\begin{equation}
\label{eq:multimom}
M_i^j = 
  \int_{V} {d}{\bf r} 
           \left[ \rho_{\rm M} +
                  \sigma_{\rm M} \right]
           \varrho^i Y_i^j(\theta,\varphi)
\;,
\end{equation}
with the coordinates centered on the lattice site. Note here that
$\varrho^i$ is the distance from the center of the cell raised to the
$i$-th power. For our cubic lattice, each quadrant of the cube is
contributed by a different region of constant $\rho_{\rm M}$ from the
dual lattice (similarly for $\sigma_{\rm M}$). With the geometry of
the lattice fixed, the multipole expansion coefficients are easily
calculated and summed to yield the total expansion coefficients for
the lattice site.

Each level of the hierarchy involves grouping cells into successively
larger cubes that completely contain cubes at the lower level. The
obvious hierarchy with each larger cube containing eight of the
smaller cubes apparently works best. In Fig.~\ref{fig:hierarchy},
cubes of the second level are separated by dashed lines and those of
the third level by solid lines. Since the number of nodes in any
direction along the simulation lattice is not restricted to a power of
two, cubes do not always contain eight smaller cubes. The $M_i^j$ for
the larger cube can be rapidly evaluated from those of the smaller
cubes using the rule for translation of a multipole expansion
\cite{GREENGARD}
\begin{equation}
\label{eq:translate}
M_i^j = \sum_{k=0}^{i} \sum_{l=-k}^{k}
  O_{i-k}^{j-l} 
  \frac{J_{l}^{j-l} A_{k}^{l} A_{i-k}^{j-l}}{A_{i}^{j}}
  \varrho^k Y_{k}^{-l}\left(\theta,\varphi\right)
\;,
\end{equation}
where $O_i^j$ are the expansion coefficients for the smaller cube and
the spherical coordinates $(\varrho,\theta,\varphi)$ here are for the
vector from the origin of the large cube to that of the small cube.
The relations for the new factors are \cite{GREENGARD}
\begin{equation}
\label{eq:define:A}
A_i^j = \frac{(-1)^j}{\sqrt{(i-j)!(i+j)!}}
\end{equation}
and 
\begin{equation}
\label{eq:define:J1}
J_{i}^{j} = 
\left\{
\begin{array}{lr}
  (-1)^{{\rm min}(|i|,|j|)} \qquad & {\rm if}\; i j < 0\\
  1 \qquad                          & {\rm otherwise}.
\end{array}
\right.
\end{equation}
The construction of the hierarchy terminates when the entire system is
enclosed by a single cube. The partitioning can be generalized to
noncubic rectangular prisms, but the restriction that the multipole
expansion is only valid at distances larger than twice the diagonal of
the rectangle will complicate the algorithm.

A downward pass through the hierarchy involving three types of
operations is required to construct $\phi_{\rm M}$ as represented by
the local expansion coefficients $L_i^j$ for each cube. The first
operation is the translation of the local expansion for the
encompassing bigger cube (if one exists) that includes all the
contributions from its far field, {\em i.e.\/}, those cubes on its
level that are not its neighbors. This translation is accomplished
by the rule \cite{GREENGARD}
\begin{equation}
\label{eq:translate_local}
L_i^j = \sum_{k=i}^{p} \sum_{l=-k}^{k}
  O_{k,l}
  \frac{\bar{J}_{k-i,l-j}^{l} A_{k-i}^{l-j} A_{i}^{j}}{A_{k}^{l}}
  \varrho^{k-i} Y_{k-1}^{l-j}(\theta,\varphi)
\;,
\end{equation}
where $O_i^j$ are the local expansion coefficients of the larger cube,
and the spherical coordinates $(\varrho,\theta,\varphi)$ here are for
the vector from the origin of the smaller cube to that of the larger
cube. $A_i^j$ is defined in Eq.~(\ref{eq:define:A}), while
\cite{GREENGARD}
\begin{equation}
\label{eq:define:J2}
\bar{J}_{i,j}^{k} = 
\left\{
\begin{array}{lr}
(-1)^i(-1)^j \qquad     & {\rm if} \; j k <0\\
(-1)^i(-1)^{k-j} \qquad & {\rm if} \; j k > 0 \, {\rm and} \, |k|<|j|\\
(-1)^i \qquad           & {\rm otherwise}.
\end{array}
\right.
\end{equation}
The next operation incorporates the contributions from areas in the
near region of the larger cube, but in the far region of the smaller
cube. This is accomplished by transforming the multipole expansion of
the source into a local expansion of the smaller cube using \cite{GREENGARD}
\begin{equation}
\label{eq:localform}
L_i^j = \sum_{k=0}^{p} \sum_{l=-k}^{k}
  O_k^l
  \frac{\tilde{J}_{k}^{j,l} A_k^l A_i^j}{A_{i+k}^{l-j}}
  \frac{Y_{i+k}^{l-j}(\theta,\varphi)}{\varrho^{i+k+1}}
\;,
\end{equation}
where the spherical coordinates are for the vector from the origin of
the smaller cube to that of the larger cube and \cite{GREENGARD}
\begin{equation}
\label{eq:define:J3}
\tilde{J}_{i}^{j,k} = 
\left\{
\begin{array}{lr}
  (-1)^i(-1)^{{\rm min}(|j|,|k|)} \qquad & {\rm if}\; j k > 0\\
  (-1)^i \qquad                          & {\rm otherwise}.
\end{array}
\right.
\end{equation}
The third type of operation is used for termination of the algorithm
at the lowest level, where the near-region contributions for the
smallest cubes must be evaluated exactly. Since a regular lattice is
used, the point ${\bf r}_i$ where $\phi_{\rm M}$ is being calculated
always lies at the corner of the neighboring cube of constant
$\rho_{\rm M}$, and the contribution is simply
\begin{equation}
\frac{\left(\Delta r\right)^2}{8}
\left(6 \,{\rm sinh}^{-1}\left(\frac{1}{\sqrt{2}}\right)
     - \frac{\pi}{2}
\right) 
\rho_{\rm M}\left({\bf r}_d\right)
\;,
\end{equation}
because only one quadrant of the region of constant $\rho_{\rm M}$
needs to be considered in this exact manner. The contribution for
squares of surface charge that touch ${\bf r}_i$ is 
\begin{equation}
\Delta r \,{\rm sinh}^{-1} \left(1\right)
\sigma_{\rm M}\left({\bf r}_d\right)
\;.
\end{equation}
Similar expressions can be calculated for general rectangular prisms.

The FMM is an efficient way of determining the magnetic potential
$\phi_{\rm M}({\bf r}_i)$ associated with a particular configuration of
a dipole field. The local observed magnetic field due to the other
dipoles is obtained using
\begin{equation}
\label{eq:hdipole}
{\bf H}_{\rm d}({\bf r}_i) = - {\bf\nabla}\phi_{\rm M}({\bf r}_i)
\end{equation}
from potential theory. Numerically, this gradient was estimated with a
centered difference of the nearest-neighbor sites, except on the
boundaries, where forward or backward differencing was used. Note that
the discretizations of all operators have been chosen for consistency
between surface and volume charges such that the model magnet can be
padded by volumes of lattice sites with ${\bf M}=0$ without changing
the results.

The FMM was implemented using {\tt C++} and named {\tt
Hierarchy.h}. Fundamental to the implementation is the class {\tt
sh\_expansion} whose instances are a $(p+1)^2$ array of complex
numbers representing either local or multipole expansion
coefficients. The class includes methods for indexing expansion
coefficients within an expansion, evaluating the expansion at a
specified point, and transforming an expansion using
Eqs.~(\ref{eq:translate}), (\ref{eq:translate_local}), and
(\ref{eq:localform}). In addition, the class encapsulates the
precomputation of quantities that do not depend on the displacement
vector, and it can return a pointer to a table of precomputations that
do depend on a fixed displacement. For instance, for efficient
evaluation of the translation of multipole expansions,
Eq.~(\ref{eq:translate}), the class contains the $(p+1)^4$ values
\begin{equation}
\label{eq:define:C1}
C_{i,k}^{j,l}=\frac{J_{l}^{j-l} A_{k}^{l} A_{i-k}^{j-l}}{A_{i}^{j}}
\;
\end{equation}
and an indirect-indexing array that specifies the sequence of
$O_{k,l}$ after the loops have been unrolled. In addition, a pointer
to the $(p+1)^2$ precomputed values of $\varrho^k
Y_{k}^{-l}(\theta,\varphi)$ can also be returned. Thus the
transformation can be evaluated efficiently and with minimal overhead
from loop-control variables.

The computer memory requirements to store the precomputations that
depend on spatial relationships can be greatly reduced for
implementations that assume a regular lattice for spatial
decomposition. For each level of the hierarchy, there will be a fixed
number of displacements between the cells in that level and with cells
on the parent level and the child level. The class {\tt sh\_expansion}
encapsulates this efficiency at each request for a precomputation by
searching a linked list of previous results and creating a new result
only when no previous result exists. A similar scheme is also
necessary for the compilation of multipole expansion coefficients for
the lowest level of the hierarchy, Eq.~(\ref{eq:multimom}). Since the
precomputations represent a significant amount of the memory
requirements of the overall simulation, these memory savings can
greatly increase the number of lattice points used to represent the
model magnet.

The partitioning of space is accomplished through the class {\tt
hbox}. Instances of this class contain geometric information about the
decomposition cell, expansions for both the potential and charges
within the cell, as well as links to other cells. These links point to
the encompassing cell, the encompassed cells, a list of same-level
cells in the far field (the ``interacting cells''), and a list of
nearest-neighbor cells. The {\tt hbox}s for each level of the
decomposition hierarchy are held in a container class {\tt hlevel},
and the entire hierarchy is maintained as a linked list. Our
implementation of {\tt hlevel} is particular to the cubic-lattice
decomposition of space, and irregular geometries have to be padded
with empty space.  (This is also true for our implementation of the
classes {\tt VectorField} and {\tt ScalarField} from {\tt
VectorField.h} used throughout our numerical integration of the LLG
equation.)

\section*{Appendix 2}

A useful way to represent the thermal Landau-Lifshitz-Gilbert
equation is in a form with the deterministic and stochastic parts
separated
\begin{equation}
\label{eq:matoem}
d{\bf M}({\bf r}_i,t) = 
  {\bf B}\left({\bf M}\left({\bf r}_i,t\right)\right)
  {\bf H}_{\rm det}\left({\bf r}_i,t\right)
  dt
+
  \sqrt{\epsilon}\,
  {\bf B}\left({\bf M}\left({\bf r}_i,t\right)\right)
  d{\bf W}\left({\bf r}_i,t\right)
\;,
\end{equation}
where ${\bf H}_{\rm det}\left({\bf r}_i,t\right)$ is the deterministic
part of the local field at ${\bf r}_i$ and ${\bf H}_n\left({\bf
r}_i\right)=\sqrt{\epsilon} d{\bf W}\left({\bf r}_i,t\right)$ is the
stochastic part. The matrix ${\bf B}$ is given by
\begin{equation}
{\bf B}
\left(
  {\bf M}\left({\bf r}_i,t\right)
\right) =
\frac{1}{1+\alpha^2}
\left(
  \begin{array}{ccc}
  \alpha \left(M_y^2+M_z^2\right) & 
  -M_z - \alpha M_x M_y           &
   M_y - \alpha M_x M_z           \\
   M_z - \alpha M_x M_y           & 
  \alpha \left(M_x^2+M_z^2\right) &
  -M_x - \alpha M_y M_z           \\
  -M_y - \alpha M_x M_z           & 
   M_x - \alpha M_y M_z           &
  \alpha \left(M_x^2+M_y^2\right) \\
  \end{array}
\right)
\;,
\end{equation}
where $x$, $y$, and $z$ represent the Cartesian coordinates and the
space and the time dependence of the $M_\mu$ have been omitted for
clarity.  The stochastic nature of the field results from the Wiener
\cite{GARD,KAMP81,KAMP} process ${\bf W}\left({\bf r}_i,t\right)$
which has the properties
\begin{equation}
\langle {W}_{\mu}\left({\bf r}_i,t\right) \rangle = 0
\qquad
\langle {W}_{\mu}\left({\bf r}_i,t\right)
        {W}_{\mu'}\left({\bf r}_i',t'\right)
\rangle 
= (t-t') \delta_{\mu,\mu'} \delta_{i,i'}
\;.
\end{equation}
The stochastic differential equation (\ref{eq:matoem}) can be
treated numerically using first-order Euler integration. 
For small $\Delta t$, the deterministic integral is
\begin{equation}
\label{eq:determin}
{\bf I}_{\rm det}\left({\bf r}_i,t\right) =
  {\bf B}\left({\bf M}\left({\bf r}_i,t\right)\right)
  {\bf H}_{\rm det}\left({\bf r}_i,t\right) \Delta t
\;.
\end{equation}
The integral of the stochastic part,
\begin{equation}
\label{eq:stodiff}
{\bf I}_{\rm sto}\left({\bf r}_i,t\right) =
  \sqrt{\epsilon}
  \int_{t}^{t+\Delta t}
  {\bf B}\left({\bf M}\left({\bf r}_i,t'\right)\right)
  d{\bf W}\left({\bf r}_i,t'\right)
\;,
\end{equation}
takes more consideration since it involves the product of the
magnetization with the Wiener process. In such cases of multiplicative
noise, different methods for evaluating Eq.~(\ref{eq:stodiff})
correspond to different Fokker-Planck equations.
\cite{GARD,KAMP81,KAMP,SANMIGUEL} There are an infinite number of
ways to interpret Eq.~(\ref{eq:stodiff}), but usually only the two
extreme cases, the It\^o and Stratonovich interpretations, are
considered. The Fokker-Planck equation considered by Brown
\cite{BROW63} only has the proper equilibrium properties when
interpreted in the Stratonovich sense. \cite{GARCIA} This is
complicated by the fact that numerical implementation of the
stochastic integral is particularly convenient in the It\^o
interpretation. Then the discretized integral is \cite{KLAUDER}
\begin{equation}
\label{eq:implement}
{\bf I}_{\rm sto} \left({\bf r}_i,t\right) =
  \sqrt{\epsilon\,\Delta t}\,
  {\bf B}\left({\bf M}\left({\bf r}_i,t\right)\right)
  {\bf g}\left({\bf r}_i,t\right)
\;,
\end{equation}
where each component of ${\bf g}$ is a random number from a Gaussian
distribution with zero mean and unit variance. Fortunately, changing
interpretations can be accomplished through a correction discussed
below. Then, Eq.~(\ref{eq:implement}) can be combined with
Eq.~(\ref{eq:determin}), and the result rearranged, to give the
``impulse'' during the integration step, Eq.~(\ref{eq:impulse}).

Changing from the Stratonovich interpretation of the stochastic
integral to that of It\^o requires the addition of a deterministic term.
Specifically, a Stratonovich interpretation of a multivariate-Langevin
equation of the form
\begin{equation}
\label{eq:stratl}
d {\bf x} = {\bf a} \left({\bf x} , t \right) dt
          + {\bf b} \left({\bf x} , t \right) d{\bf W}
\;,
\end{equation}
is equivalent to the Langevin equation \cite{GARD}
\begin{equation}
\label{eq:itol}
d {\bf x} = \left( {\bf a} \left( {\bf x} , t \right)
                 + {\bf c} \left( {\bf x} , t \right) \right) dt
          + {\bf b} \left(  {\bf x} , t \right) d{\bf W}
\end{equation}
in the It\^o interpretation \cite{GARD,KAMP} where the components of
the new drift term are
\begin{equation}
 c_\mu = \frac{1}{2} 
         \sum_{\nu,\nu'} b_{\nu',\nu} 
                         \frac{\partial b_{\mu,\nu}}
                              {\partial \nu'}
\;.
\end{equation}
For the present system, the additional drift term is readily found to
be
\begin{equation}
{\bf c}({\bf r}_i,t) = \frac{\epsilon}{(1+\alpha^2)}
                       {\bf M}({\bf r}_i,t)
\;,
\end{equation}
which is equivalent to the result given in Ref.~\onlinecite{GARCIA}.  This
drift term is always directed along the local magnetization density.
The process of normalizing the magnitude of the spins during each
integration step, Eq.~(\ref{eq:mnew}), is also directed along the
magnetization, and essentially takes the correction into account.

\vskip 1in

\center{
\begin{tabular}{|l|l|}
\hline
$\ell_e$		& $3.6\,{\rm nm}$ 			\\
$M_s$			& $1700\,{\rm emu/cm^3}$  		\\
$\Delta{\bf\tilde{r}}$	& $1.5\,{\rm nm}$			\\
$\Delta \tilde{t}$	& $5.0 \times 10^{-5}\,{\rm ns}$	\\
$\gamma_0$		& $1.76 \times 10^{7}\,{\rm Hz/Oe}$	\\
$\alpha$		& $0.1$					\\
\hline
\end{tabular}
}

\noindent
{Table 1. Parameters used in the micromagnetic simulations
described here.}

\newpage

~

\vskip .25in
~

\newpage
~

~

\newpage
~

~

\newpage
~

\newpage
~

~

\newpage
~

~


\begin{references}
\bibitem{EXCHANGE} The value of the exchange length used, $l_{\rm
e}$$=$$3.6\,{\rm nm},$ here has been adjusted to account for a
multiplicative factor of $2$ that we initially believed should appear
in Eq.~(3), see Ref.~\protect{\onlinecite{BandB}}. A more common value
for iron is $l_e=2.6\,{\rm nm}$ (see Refs.~\protect{\onlinecite{CULLITY}}
and \protect{\onlinecite{WIRTH98}}), which is approximately $1/\sqrt{2}$
times the value used here.
\end{references}
\end{document}